\documentclass[12pt]{article}
\usepackage{amssymb,amsmath}
\usepackage{graphicx}
\date{}
\title{Long quantum transitions due to unstable semiclassical
  dynamics} 
\author{D.G.~Levkov\footnote{levkov@ms2.inr.ac.ru}, 
        A.G.~Panin\footnote{panin@ms2.inr.ac.ru}\\
{\small  Institute for Nuclear Research of the Russian Academy of
  Sciences,}\\ {\small 60th October Anniversary Prospect 7a, Moscow
  117312, Russia}
}
\begin{document}
\maketitle
\begin{abstract}
Quantum transitions are described semiclassically as motions of
systems along (complex) trajectories. We consider the cases when the
semiclassical trajectories are unstable and find that durations of
the corresponding transitions are large. In addition, we show that 
the probability distributions over transition times have unusual
asymmetric form in cases of unstable trajectories. We investigate in
detail  three types of processes related to unstable semiclassical
dynamics. First, we analyze recently proposed mechanism of
multidimensional tunneling where transitions proceed by formation and
subsequent decay of classically unstable ``states.'' The second class
of processes includes one--dimensional activation transitions due to
energy dispersion. In this case the semiclassical transition--time
distributions have universal form. Third, we investigate long--time
asymptotics of transition--time distributions in the case of
over--barrier wave packet transmissions. We show that behavior of
these asymptotics is controlled by unstable semiclassical trajectories
which linger near the barrier top.
\end{abstract}

\section{Introduction}
\label{sec:introduction}
Semiclassical method unveils fascinating relation between the
properties of quantum transitions and the character of classical
dynamics of the system. This relation is valid, in particular, for
tunneling processes~\cite{Creagh:1998} which do not exist at the
classical level. Indeed, tunnelings are described semiclassically by
{\it complex trajectories}~\cite{Miller:1974} evolving classically in
complex phase 
space. Observables characterizing tunneling processes, such as
probabilities~\cite{Miller:1974} or splittings of 
energy levels~\cite{Takada1994}, are functionals of complex
trajectories. Clearly, features of (complex) classical dynamics
are imprinted in these observables.

The link between tunneling processes and classical dynamics is
supported by theoretical studies of multidimensional tunneling. It was
found that behavior of tunnel splittings of energy levels is
qualitatively different for systems with integrable~\cite{Miller:1968},
near--integrable~\cite{Wilkinson:1986}, mixed~\cite{Bohigas:1993} and
chaotic~\cite{Creagh:1999} dynamics. Some of the related phenomena
have already been observed experimentally~\cite{Dembowski:2000}. 

In this paper we study impact of non--trivial semiclassical
dynamics on temporal characteristics of classically forbidden
transitions. We are particularly interested in cases when complex
trajectories describing transitions are unstable. Such trajectories do
not interpolate directly between in-- and out-- regions of the
process, but rather get attracted to a certain unstable orbit lying on
the boundary between the regions. It takes large or even infinite
time for a trajectory of this kind to arrive to the
out--region. In general one does not expect the time scale of the
semiclassical motion to coincide with the (properly defined) 
duration of quantum transition. Nevertheless, we advocate qualitative
relation~\cite{Takahashi:2006} between these quantities: 
durations of quantum processes are large when the corresponding
semiclassical trajectories are unstable.

\begin{figure}
\centerline{\includegraphics[width=0.7\textwidth]{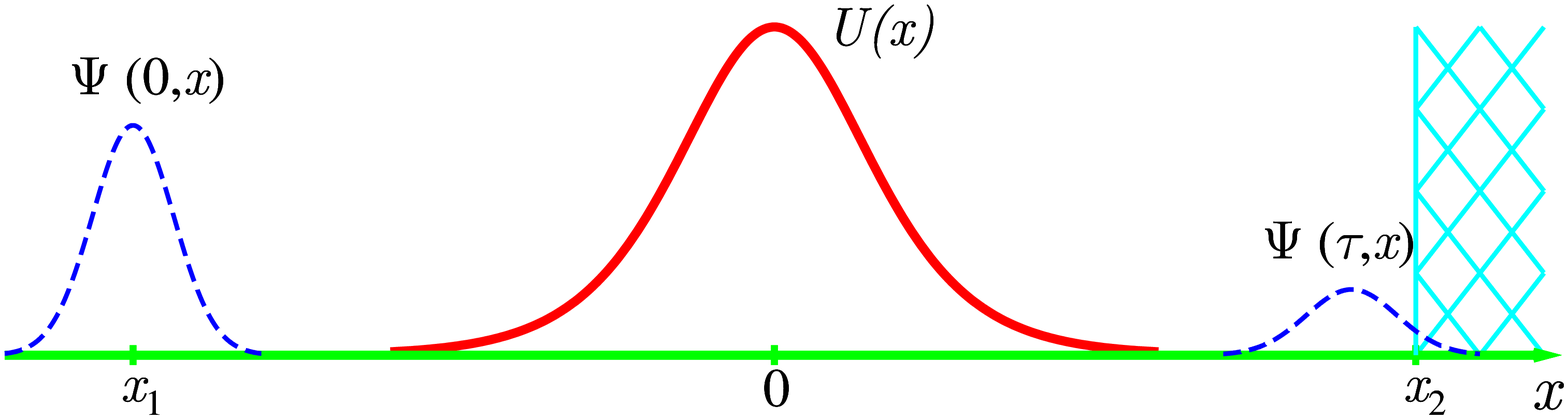}}
\caption{\label{fig:1}(Color online) Experimental setup for measuring 
  the traversal--time distribution $\rho(\tau)$. Only the ``transition''
  coordinate $x$ is shown. Macroscopic detector is represented by
  the hatched region $x>x_2$.}
\end{figure}

Presently there is no generally accepted
definition of quantum traversal time~\cite{timereviews}. During 
last decades a number of definitions have been suggested in
literature~\cite{Bohm:1951,Smith:1959,Baz:1967,Buttiker:1982,
Sokolovski:2008}. Below we follow the approach of 
Refs.~\cite{Ohmura:1964,Jaworski:1987,Olkhovsky:2001} where
traversal time $\tau$ is treated as a random variable with
probability distribution $\rho(\tau)$. Experiment for
measuring $\rho(\tau)$ is depicted schematically in
Fig.~\ref{fig:1}. The horizontal 
axis in this figure represents ``transition'' coordinate $x$ which
runs between in-- and out-- regions of the process and meets potential
barrier $U({\bf x})$ on the way. Quantum particle is emitted at $t=0$,
$x\approx x_1$ in a definite initial state $\Psi_i({\bf x}) =
\Psi(0,{\bf x})$. As  time goes on, the particle moves
towards the potential barrier and eventually traverses it. At
$t=\tau$ the particle is caught by the asymptotically distant
macroscopic detector which occupies the region $x>x_2$ and registers
the time of arrival $\tau$. The probability of registration is
\begin{equation}
\label{eq:55}
{\cal P}_\tau = \int\limits_{x>x_2} d{\bf x}_f\; |\Psi(\tau,{\bf
  x}_f)|^2\;,
\end{equation}
where the integration variable is denoted by ${\bf x}_f$ for later
convenience. 
Note that the total transmission probability ${\cal P}_{\infty}$ is
obtained from ${\cal P}_\tau$ in the limit $\tau\to
+\infty$. We also remark that the points $x_1$ and $x_2$ are assumed
to be far enough from the potential barrier, so that the particle
moves freely at $x\lesssim x_1$, $x \gtrsim x_2$.
One introduces the normalized probability density
\begin{equation}
\label{eq:1}
\rho(\tau) = \frac{1}{{\cal P_{\infty}}} \frac{d{\cal
    P}_\tau}{d\tau}\;.
\end{equation}
By construction, $\rho(\tau)\cdot d\tau$ is the increment of the
registration probability, which is conditional with respect to the
transition event. Following\footnote{There
  are two important simplifications in our setup. First, we prepare
  definite in--states of the process. Second, we consider the case of 
  asymptotically distant detector, which saves us from the
  controversial issue of separating transmitted and reflected parts of
  wave function.}
Refs.~\cite{Ohmura:1964,Jaworski:1987,Olkhovsky:2001},
we associate $\rho(\tau)$ with probability distribution over
traversal times $\tau$. The average time of passing is then computed
as 
\begin{equation}
\label{eq:15}
\langle \tau \rangle = \int_0^{+\infty} d\tau \;\tau \rho
(\tau) \;,
\end{equation}
and time dispersion $\sigma_\tau^2$ is defined
accordingly.

It is worth noting that several assumptions about detector registering
the particles have been made in Eq.~(\ref{eq:55}). First, we adopted
approximation of instantaneous measurements disregarding time
resolution of the detector. Second, we assumed that the states of
particles are not distorted prior to the measurement 
event. Third, our probability formula  (\ref{eq:55}) is applicable
only in the case of small registration probabilities: true probability
of registering the particle at $t=\tau$ is {\it conditional} with
respect to previous non--detection. For classically forbidden
transitions considered below the probability of non--detection is
close to $1$ and  Eq.~(\ref{eq:55}) is applicable. In other cases
expression  for ${\cal P}_\tau$ should be corrected\footnote{One can 
  simply consider detectors registering particles with probabilities
  $\nu {\cal P_\tau}$, where $\nu$ is a small number. In this case
  Eq.~(\ref{eq:55}) is valid, since the probabilities of previous
  detections are small.}~\cite{Anast:2006}.

Equipped with the definition (\ref{eq:15}), we study three examples
of quantum processes related to unstable semiclassical
trajectories. In all cases our results confirm the qualitative
conclusion~\cite{Takahashi:2006} about large transition
times. Namely, we show that $\langle \tau \rangle \sim |\log g^2|$,
$\sigma_\tau^2 \gtrsim O(g^0)$, where $g^2$ is the semiclassical
parameter\footnote{This is a dimensionless   quantity   proportional
  to $\hbar$ which measures the relative size of quantum
  fluctuations. Semiclassical expressions become accurate 
  as $g^2\to 0$.}. In the limit of vanishingly small
$g^2$ when the  system becomes increasingly ``more classical'' the
average duration of transition grows to infinity, while
$\sigma_\tau^2$ stays finite or grows. This behavior is drastically
different from the respective scalings in the case of stable
trajectories where $\langle \tau \rangle$ reaches definite limit and
$\sigma_\tau^2$ vanishes as $g^2\to 0$.

We consider the following classically forbidden processes. First and
most importantly, we 
study tunneling transitions in non--linear multidimensional systems at
high energies. Instability of semiclassical dynamics in this case was
demonstrated in Refs.~\cite{Bezrukov:2003yf,Takahashi:Ikeda}, see
Refs.~\cite{Onishi:2003} for earlier works. We have 
already sketched the behavior of the corresponding complex
trajectories: they approach certain unstable orbit, remain close to it
for an infinite time interval and then slide away into the
out--region. In systems with two degrees of freedom the mediator
unstable orbit describes periodic oscillations near the saddle point
of the potential. We call 
it {\it sphaleron}\footnote{This term is taken from field
  theory~\cite{Klinkhamer:1984di}. We find it more convenient than 
  exact but lengthy expression ``normally hyperbolic invariant
  manifold'' (NHIM)~\cite{Wiggins:2001}.} or simply {\it unstable
  periodic orbit}. We also attribute instability of
respective trajectories to appearance of new mechanism of {\it
  sphaleron--driven} tunneling (or tunneling {\it driven by unstable
  periodic orbit}). Recent
studies~\cite{Bezrukov:2003er,epsilon,Takahashi:2008} revealed
two experimental signatures of this mechanism: 
the probabilities of related processes are suppressed by 
additional power--law factor~\cite{epsilon} and 
distributions over out--state quantum numbers are anomalously 
wide~\cite{Takahashi:Ikeda,epsilon,Takahashi:2008}. Below we show that
the distribution $\rho(\tau)$ and corresponding 
values of $\langle \tau \rangle$ and $\sigma_\tau^2$ can be used
for experimental identification of the sphaleron--driven 
mechanism of tunneling. 

Second, we investigate wave packet transmissions through
one--dimensional 
potential barrier. We consider the case when  the average energy of
wave packet is lower than the height of the barrier but sufficiently
close to it. Then transmission proceeds by {\it activation} caused by
energy uncertainty in the initial state. The probability of such
transmission is exponentially suppressed, while the corresponding
semiclassical trajectory approaches the top of the barrier and stays
there for an infinite time. This trajectory is unstable; its behavior
is, again, of the type described above\footnote{The constituent
  unstable orbit in this case is static classical solution ``sitting''
  on top of the barrier.}. Our calculations show that the forms of the
semiclassical traversal--time distributions are {\it universal} in one
dimension: all information about the potential barrier and initial
wave packet is encoded in two parameters of the distributions, which
can be absorbed by shifting and rescaling $\tau$. Universal forms of
$\rho(\tau)$ lead to universal qualitative dependencies $\langle
\tau \rangle \sim |\log g^2|$, $\sigma_\tau^2 \sim O(g^0)$ in the case
of one--dimensional activation transitions.

Third and finally, we study the probabilities of large time delays
$\tau$ in one--dimensional over--barrier transmissions. In this case
the average energy of particle exceeds the height of the
potential barrier. The particle is registered at time $\tau$ after
transmission at a given distance behind the
barrier; we calculate semiclassically the asymptotics $\tau\to
+\infty$ of the registration probability. Semiclassical
trajectories entering this calculation linger near the barrier
top. The probability at large $\tau$ has  universal form, again. Due
to delays of the respective trajectories this probability is larger than
one would expect naively. 

We illustrate our findings by performing explicit 
calculations in systems with one and two degrees of freedom. 
In these systems we compare the semiclassical results for $\rho(\tau)$
with results of explicit quantum--mechanical calculations. We find
agreement at small $g^2$. 

The paper is organized as follows. We start in
Sec.~\ref{sec:tunn-time-defin} with properties of
traversal--time definition (\ref{eq:15}). Then, in
Sec.~\ref{sec:sphal-driv-tunn} we discuss the  mechanism of
sphaleron--driven tunneling and one--dimensional activation
processes. General semiclassical expression for $\rho(\tau)$ is
introduced in Sec.~\ref{sec:semiclassical-method-1}. Results for
the durations of quantum processes are presented in
Sec.~\ref{sec:results}: we consider one--dimensional activation
transitions in Sec.~\ref{sec:one-dimension-1}, investigate
sphaleron--driven tunneling in Sec.~\ref{sec:two-dimensions-1} and 
study long--time asymptotics of ``over--barrier'' traversal--time
distributions in Sec.~\ref{sec:long-time-asympt}. We summarize in
Sec.~\ref{sec:concluding-remarks}. Details of semiclassical
and exact quantum--mechanical calculations are presented in
Appendices.

\section{Properties of traversal--time definition}
\label{sec:tunn-time-defin}
The controversial issue of quantum traversal time has attracted
considerable attention during last decades~\cite{timereviews}. The 
reason for the controversy seems to be hidden within the quantum theory
itself which does not offer clear definition or unique method 
of computing the durations of quantum transitions. At the moment there
exists a number of physically different definitions of quantum 
traversal time, see e.g. Refs.~\cite{Bohm:1951,Smith:1959, 
Baz:1967,Buttiker:1982,Sokolovski:2008}. 

Expression (\ref{eq:15}) belongs to a wide class of 
definitions involving delay of the transmitted wave packet with
respect to the incident one. In this approach the delay is computed
using some feature of wave packet, maximum~\cite{Bohm:1951} or 
centroid~\cite{Leavens:1989}. In our case the probability  
conservation law implies that the distribution~(\ref{eq:1}) is
proportional to the total current through the surface $x=x_2$ at
$t=\tau$. Thus,  $\rho(\tau)$ reflects the form of the transmitted
wave packet, and mean time of passing (\ref{eq:15}) fits into the
above class of wave packet--related definitions. We remark that
$\rho(\tau)  \cdot d\tau$ can be regarded as a  probability of
registering the particle at time $\tau$ by the point--like detector
situated at $x = x_2$.

The properties of the definition (\ref{eq:15}) are similar to those of
other definitions from the same class. First, the setup used in the
Introduction resembles recent experiments with
photons~\cite{Steinberg:1993,Winful:2006}, and one can hope to 
measure the distribution $\rho(\tau)$ experimentally. Note, however, 
that experimental verification of our results, if possible, requires
substantial modification of existing experiments. Indeed, recent
measurements of traversal times have been 
performed in quasi--stationary regime where wave functions of
particles are substantially wider than potential barriers. In this
case the  interval (\ref{eq:15})
coincides~\cite{Leavens:1989,Olkhovsky:2001} with the phase time of 
Bohm and Wigner~\cite{Bohm:1951} which is reproduced in actual
measurements~\cite{Steinberg:1993}. On the other hand, our results are
obtained in the ``semiclassical'' case where the coordinate
uncertainty is parametrically smaller than the barrier width.

Second, the interval~(\ref{eq:15}) passes~\cite{Olkhovsky:2001}
self--consistency check of Refs.~\cite{timereviews}: times of
transmission through the barrier and reflection from it have sense of
conditional averages over two mutually exclusive events. 

Third, our traversal time (\ref{eq:15}) suffers from 
Hartman effect~\cite{Hartman:1962} which leads to superluminal
velocities of under--barrier motions. This effect was 
observed experimentally~\cite{Winful:2006}. It is explained as
follows. One notices~\cite{timereviews,Japha:1996} that the out--state
of the tunneling process is formed by the waves constituting
forward tail of the initial wave packet. Thus, the out--state is
not related casually to the central part of the in--state wave
function. On the other hand, only the central parts of in-- and out--
states contribute 
substantially into  $\langle \tau \rangle$. Since casual connection
between these parts is absent, apparent velocity
$(x_2-x_1)/\langle \tau \rangle$ can be superluminal. We stress that
the definition 
(\ref{eq:15}) is sufficient for the  purposes of this paper:
we regard $\langle\tau \rangle$ as an experimentally
measurable~\cite{Winful:2006} quantity characterizing quantum
transitions.

Finally, let us remark that quantum traversal time can be naturally
defined without any reference to the delay between transmitted and
incident wave packets: one can exploit stopwatches
attached to the particle~\cite{Baz:1967} or observe response of
total transmission probability to the periodic modulation of the
potential~\cite{Buttiker:1982}. In these approaches the properties of
traversal times are essentially different from those listed
above. Although we believe that the qualitative  
conclusions of this paper should be valid for any reasonable setup,
our quantitative results are not applicable for other traversal--time
definitions.

\section{Three processes}
\label{sec:sphal-driv-tunn}
Let us describe three types of processes
related to unstable semiclassical dynamics. 

We start with one--dimensional activation transitions 
of quantum particles through potential barriers
$U(x)$. In what follows we consider the 
setup in Fig.~\ref{fig:1}. The initial state $\Psi_i(x)$ 
is chosen to be a Gaussian wave packet with central
position $x_1$, average momentum $p_0$ and momentum dispersion
$\sigma_p^2$. Transmissions proceed between the in-- and out--
regions, $x\to -\infty$ and $x\to +\infty$ respectively. Points $x_1$
and $x_2$ in Fig.~\ref{fig:1} belong to these regions.

We exploit the semiclassical approximation which is
justified if
\begin{equation}
\label{eq:27}
g^2 \equiv \frac{\hbar}{l_0 \sqrt{mV_0}}\ll 1\;.
\end{equation}
Here $V_0$ and $l_0$ are height and width of the potential barrier
and $m$ is the particle mass. Dimensionless parameter
$g^2$ introduced in Eq.~(\ref{eq:27}) measures accuracy of 
the semiclassical approximation; one can adjust the value of this
parameter by changing $U(x)$. Below we adopt
dimensionless units $m = \hbar = V_0/l_0^2 = 1$. In these units the
height $V_0 = 1/g^2$ and width $l_0 = 1/g$ of the ``semiclassical''
barrier are parametrically large.

We use the following scalings of the in--state parameters with
$g^2$. First, we suppose 
that the typical energy of the process is large,
$p_0^2 \sim V_0 \sim 1/g^2$. Second, momentum dispersion is chosen to  
be of order one, $\sigma_p^2 \sim O(g^0)$. Finally, the points $x_1$ and $x_2$
are taken to be sufficiently far from the potential barrier, so that
motion is free at $x\lesssim x_1$, $x\gtrsim
x_2$. This implies $x_1,\, x_2 \sim 1/g$.

We have already mentioned in the Introduction that one--dimensional
transmissions proceed by activation at some values of in--state
parameters. Let us explain the origin of the effect  by 
decomposing the initial state $\Psi_i$ in the basis of plane waves
with fixed momenta $p$. We consider the case $p_0 < p_{c,2} \equiv 
\sqrt{2V_0}$ when the initial wave packet is mostly formed by
``under--barrier'' modes with $p\sim p_0<\sqrt{2V_0}$. Clearly, these
modes are exponentially suppressed after
transmission. On the other hand, ``over--barrier'' modes with
$p>\sqrt{2V_0}$ constitute exponentially small fraction of the
initial state, but overcome the barrier with probability of order
one. One sees that, depending on the  suppressions, the probability 
${\cal P}_\infty$ can be dominated  by
``under--barrier'' or ``over--barrier'' modes. We call the
respective transitions by tunneling and activation. Note that the
value of ${\cal P}_\infty$ is exponentially suppressed in both cases. 
Note also that the
activation processes are related to the possibility of
overcoming the barrier due 
to finite momentum dispersion in the in--state. 

In Sec.~\ref{sec:one-dimension-1} we will argue that there always
exists some critical value of momentum $p_{c,1}$ such that
transitions proceed by activation in the region $p_{c,1}
< p_0 < p_{c,2}$. We will also demonstrate 
explicitly that activation transitions are described by unstable
semiclassical trajectories which linger near the barrier top. 

\begin{figure}[htb]
\centerline{\includegraphics[width=0.8\textwidth]{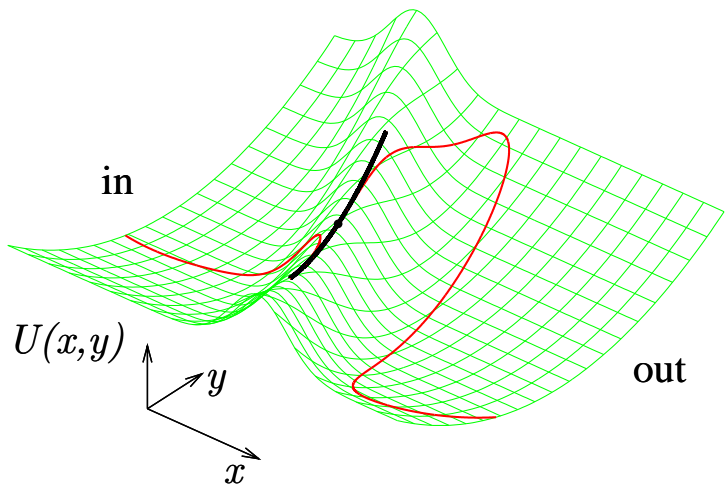}}
\caption{\label{fig:9}(Color online) Potential (\ref{eq:54}) (not to
  scale). On top of the potential surface: real part of complex
  trajectory (thin red line) describing sphaleron--driven tunneling at
  $gp_0=1.55$, $\sigma_p^2 = 0$, $g^2E_y = 0.05$ 
  and the sphaleron orbit (thick black line). The saddle point of the
  potential is shown with thick black dot.}
\end{figure}
The second type of processes is related to the multidimensional
mechanism of sphaleron--driven tunneling. For definiteness we explain 
this mechanism in the model of
Refs.~\cite{Bonini:1999kj,Bezrukov:2003yf,epsilon} where
two--dimensional quantum particle of mass $m$ moves in harmonic
potential valley of frequency $\omega_0$. The valley is intersected
at an angle by the potential barrier. The potential of the model is
\begin{equation}
\label{eq:54}
U(x,y) = \frac{m\omega_0^2}{2} y^2 + V_0\,\mathrm{e}^{-
  (x+y)^2/2l_0^2}\;,
\end{equation}
where $V_0$ and $l_0$ are height and width of the barrier,
respectively. Function $U(x,y)$ is plotted in Fig.~\ref{fig:9}. 

Semiclassical approximation in the model~(\ref{eq:54}) is
justified under conditions (\ref{eq:27}) and
\begin{equation}
\label{eq:56}
\omega \equiv \omega_0 \,l_0 \sqrt{m/V_0} \sim O(g^0)\;.
\end{equation}
In dimensionless units introduced above the
potential~(\ref{eq:54}) contains only  two parameters, $g^2$ and
$\omega$,
\begin{equation}
\label{eq:58}
U(x,y) = \omega^2 y^2/2 + \frac{1}{g^2}\,\mathrm{e}^{- g^2(x+y)^2/2}\;.
\end{equation}
We use $\omega=1/2$ for the waveguide frequency.

In what follows we study tunneling transitions between the
asymptotic regions $x\to -\infty$  and $x\to +\infty$ 
(in-- and out-- regions respectively). In--states $\Psi_i$ of
transitions are fixed completely by taking $\Psi_i(x,y) =\psi_x(x) 
\cdot \psi_y(y)$, where $\psi_x(x)$ is a Gaussian wave packet with
central position $x_1$, average momentum $p_0$ and momentum dispersion
$\sigma_p^2$, and $\psi_y(y)$ 
is an oscillator eigenfunction with energy $E_y$. We propagate
semiclassically the in--state into the out--region and
compute the probability ${\cal P}_\infty$ and
traversal--time distribution $\rho(\tau)$.

In the semiclassical regime the typical energies of motion in $y$
direction are parametrically large, $E_y \sim 1/g^2$. The
scalings of other parameters are chosen in the same way as in the
above case of one--dimensional transitions. We characterize
transitions with  the combinations $gp_0$, $g^2 E_y$, $\sigma_p^2$,
$gx_1$, $gx_2$ which stay finite as $g^2\to 0$.

Note that transmissions between the asymptotic regions of the
potential (\ref{eq:58}) remain classically forbidden even 
when the  total energy $p_0^2/2 + E_y$ exceeds the minimum height $V_0
= 1/g^2$ of the potential barrier. Indeed, classical trajectories with
energies slightly higher than $V_0$ can overcome the barrier only by
passing in  the immediate vicinity of the saddle point of the
potential. General trajectory misses this vicinity, and the
corresponding transition remains classically forbidden. The
probability of such transition is exponentially suppressed on general
grounds. Tunneling transitions at energies exceeding the height of the
potential barrier are known as {\it dynamical
  tunneling}~\cite{Miller:1974,Heller:1981}. 

The mechanism of sphaleron--driven tunneling is
based~\cite{Bezrukov:2003yf,Takahashi:Ikeda} on behavior of
complex trajectories in the regime of dynamical
tunneling. One of such trajectories is depicted in Fig.~\ref{fig:9}
(thin red line). It overcomes the barrier in two stages,
by approaching the unstable periodic orbit (sphaleron) and departing
from it. Note that the sphaleron is real; it is plotted in
Fig.~\ref{fig:9} by thick black line. This orbit is unstable by
construction, since it describes  oscillations around the saddle point
$x=y=0$ of the potential. The semiclassical trajectories in the new
tunneling mechanism remain close to the sphaleron for an infinite time
interval; imaginary parts of these trajectories become negligibly
small after several sphaleron oscillations.

In general one finds that the sphaleron--driven mechanism is relevant
in some region of in--state parameters, $p_{c,1} < p_0 < p_{c,2}$,
where $p_{c,1}$, $p_{c,2}$ 
are  critical momenta depending on $E_y$ and
$\sigma_p^2$. Regions of low and high momenta, $p_0<p_{c,1}$ and $p_0>
p_{c,2}$ respectively,  correspond to direct tunneling through the
barrier and classical over--barrier transitions. 

We remark that multidimensional processes of sphaleron--driven
tunneling are essentially different from the one--dimensional
activation transitions considered above: instability of trajectories
in the sphaleron--driven case is not caused by finite momentum
dispersion.

The third effect considered in this paper is related to the
possibility of large time delays in over--barrier wave packet
transmissions. For simplicity we return to one--dimensional setup
where Gaussian wave packets traverse the potential barrier
$U(x)$. Now, we are interested in the case of high initial momenta,
$p_0 > \sqrt{2V_0}$, when the central part of wave packet overcomes
the barrier classically. We study the tail of $\rho(\tau)$
corresponding to large traversal times $\tau$. In
Sec.~\ref{sec:long-time-asympt} we will argue that this tail can be
readily found by considering unstable semiclassical trajectories
which linger near the barrier top.  

From the physical viewpoint the origin of the effect is explained,
again, by 
decomposing the initial state into a collection of plane waves with fixed
asymptotic momenta $p$. In general, waves with $p\approx \sqrt{2V_0}$
are populated, although the respective probability is exponentially
small. These waves get delayed near the top of the barrier
and create long tail of $\rho(\tau)$.

\section{Semiclassical method}
\label{sec:semiclassical-method-1}
Here we describe the standard method of complex
trajectories~\cite{Miller:1974} and introduce convenient semiclassical
expression for $\rho(\tau)$. Although the semiclassical techniques of
this Section are pretty general, we concentrate on the one-- and
two--dimensional models of Sec.~\ref{sec:sphal-driv-tunn}.

For a start, let us briefly explain the origin of complex
trajectories~\cite{Miller:1974} leaving details of semiclassical 
computations to 
Appendix~\ref{sec:standard-method}. Consider evaluation
of total transmission probability ${\cal P}_\infty$. One starts with
the path integral for the final state,
\begin{equation}
\label{eq:60}
\Psi(\tau,{\bf x}_f) = \int d{\bf x}_i \, \Psi_i({\bf x}_i)\int [d{\bf
  x}]\Big|_{{\bf x}_i}^{{\bf x}_f} \, \mathrm{e}^{iS[{\bf x}]} \;,
\end{equation}
where the paths ${\bf x}(t)$ are defined in the time interval $t\in
[0;\, \tau]$: they start at ${\bf x}(0) = {\bf x}_i$ and
arrive to ${\bf x}(\tau) = {\bf x}_f$. $S[{\bf x}]$ is the action in
this time interval. Since we are interested in ${\cal P}_\infty$, the
value of $\tau$ will be sent to infinity in the end of calculation.

At small $g^2$ the integral
(\ref{eq:60}) can be evaluated by the saddle--point method. Complex
trajectory ${\bf x}(t)$ describing 
tunneling process is the saddle--point path of integration. In
Appendix~\ref{sec:standard-method} we obtain saddle--point equations
for this trajectory. Namely, ${\bf x}(t)$ should satisfy the classical 
equations of motion. Conditions at $t=0$ come from the saddle--point
integration over ${\bf x}_i$,
\begin{equation}
\label{eq:18}
\dot{x}_i - 2i\sigma_p^2 (x_i - x_1) = p_0\;, \qquad \qquad
E_y = (\dot{y}^2_i + \omega^2 y^2_i)/2\;,
\end{equation}
where the subscript $i$ denotes quantities computed at $t=0$. [In
the one--dimensional model of Sec.~\ref{sec:sphal-driv-tunn} only
the first condition should be imposed.] One sees that the second of
Eqs.~(\ref{eq:18}) fixes initial energy of
$y$--oscillator. Interpretation of the first equation is clear if the
trajectory ${\bf x}(t)$ is real. In this case the real and imaginary
parts of the first of Eqs.~(\ref{eq:18}) constitute two independent
conditions specifying the initial  
particle position $x_i=x_1$ and momentum $\dot{x}_i = p_0$. In the
complex case the two parts are combined with the coefficient
proportional to the momentum dispersion~$\sigma_p^2$. Final
boundary conditions for ${\bf x}(t)$ arise when one substitutes
$\Psi(\tau,{\bf x}_f)$ into the probability expression (\ref{eq:55})
and evaluates the saddle--point integral over ${\bf x}_f$. One obtains
conditions
\begin{equation}
\label{eq:19}
 \mathrm{Im}\, \dot{\bf x}(t),\; \mathrm{Im}\, {\bf x}(t) \to
 0 \qquad \qquad\mbox{as} \qquad t\to +\infty
\end{equation}
which mean that evolution in the out--region is real. Note that the
limit $\tau\to +\infty$ is already taken in Eq.~(\ref{eq:19}); at finite
$\tau$ the integral over ${\bf x}_f$ cannot be evaluated by the
saddle--point method.

After finding the trajectory ${\bf x}(t)$ from the classical
equations of motion and boundary conditions (\ref{eq:18}),
(\ref{eq:19}), one computes the transmission probability by the
semiclassical formula\footnote{We
  assume that only one trajectory 
  gives substantial contribution into the probability.}
\begin{equation}
\label{eq:59}
{\cal P}_\infty = A \cdot \mathrm{e}^{-F/g^2}\;,
\end{equation}
where the suppression exponent $F$ and prefactor $A$ are 
functionals of the saddle--point trajectory ${\bf x}(t)$. Explicit
expressions for these functionals are 
derived in Appendices~\ref{sec:standard-method} and~\ref{sec:1d},
see Eqs.~(\ref{eq:65}), (\ref{eq:67}) and (\ref{eq:21}).

Now, let us evaluate semiclassically the distribution
$\rho(\tau)$. Following straightforward method~\cite{Takahashi:2006},
one should repeat the above considerations in the case of finite
durations of transitions $\tau$ and apply Eqs.~(\ref{eq:55}),
(\ref{eq:1}). It is simpler, however, to use indirect technique
introduced in Refs.~\cite{epsilon}.

With the purpose of explaining the logic of Refs.~\cite{epsilon}, we
start from purely classical case when the traversal time $\tau$ is
defined as the time of motion in the region $x<x_2$. [Recall that real
classical trajectories start at $x=x_1$, $t=0$.] One introduces the
functional
\begin{equation}
\label{eq:2}
T_{int}[{\bf x}] = \int_0^{\infty} dt \,\theta(x_2 - x(t))\;,
\end{equation}
which returns traversal time for any classical path ${\bf x}(t) \in
\mathbb{R}$. We call $T_{int}$ {\it interaction time}. To fix the
value of $T_{int}$, one introduces Lagrange multiplier 
$\tilde{\epsilon}$ and changes the classical action, $S[{\bf x}]\to
S[{\bf x}] + \tilde{\epsilon} (T_{int}[{\bf x}]-\tau)$. Trajectories
${\bf x}_{\tilde{\epsilon}}(t)$ extremizing the new action correspond
to fixed durations $\tau = T_{int}[{\bf x}_{\tilde{\epsilon}}]$ of
transitions. Note that the Lagrange method enables one to replace
explicit fixation of traversal time $\tau$ with modification of
classical equations of motion.

It is not trivial to introduce Lagrange method in quantum
case. Indeed, there is no a priori reason to associate stochastic
variable $\tau$ with values taken by the functional
$T_{int}$. Still, we find that there exists close connection between
these 
quantities. The technical trick revealing this connection is
explained as follows. One inserts into the integrand of
Eq.~(\ref{eq:60}) the unity factor
\begin{equation}
\label{eq:7}
1 = \int_0^{\tau} d\tau' \, \delta(T_{int}[{\bf x}] - \tau') =
\int_0^{\tau} d\tau' \int_{i\infty}^{-i\infty}
\frac{id\epsilon}{2\pi g^2} \; \mathrm{e}^{\epsilon(\tau' -
  T_{int}[{\bf x}])/g^2}\;.
\end{equation}
 Since the paths 
${\bf x}(t)$ in Eq.~(\ref{eq:60}) are defined in the interval
$t\in [0;\, \tau]$, we temporarily restrict integration in
Eq.~(\ref{eq:2}) to this interval. Now, $T_{int}[{\bf x}]$ takes
values between $0$  and $\tau$; we explicitly used this fact in
the first equality of Eq.~(\ref{eq:7}). In the second equality we 
performed Fourier transformation and introduced ``Lagrange
multiplier'' $\epsilon$. The path integral for $\Psi(\tau,{\bf x}_f)$
takes the form,   
\begin{equation}
\label{eq:28}
\Psi(\tau,{\bf x}_f) = \int_0^\tau d\tau' \int_{i\infty}^{-i\infty}
\frac{id\epsilon}{2\pi g^2} 
\left \{ 
\int d{\bf
  x}_i \, \Psi_i({\bf x}_i)\int [d{\bf 
  x}]\Big|_{{\bf x}_i}^{{\bf x}_f} \, \mathrm{e}^{iS[{\bf
    x}] - \epsilon (T_{int}[{\bf x}]-\tau')/g^2}\right\} \;.
\end{equation}
Recall that the paths ${\bf x}(t)$ in this integral are defined in the
interval $t\in [0;\, \tau]$. Expression~(\ref{eq:28}) is remarkable in
three respects. First, its integrand depends on ${\bf x}(t)$ via the
new action $S[{\bf x}] + 
i\epsilon (T_{int}[{\bf x}]-\tau')/g^2$ which is similar to the action
arising in the classical Lagrange method. This is natural, since by
inserting the $\delta$--function, Eq.~(\ref{eq:7}), we introduced
explicit constraint $T_{int}[{\bf x}] = \tau'$ in the path integral. 
Second, apart from the modification of the action the integral in
brackets in Eq.~(\ref{eq:28}) has absolutely the same form as
Eq.~(\ref{eq:60}). Thus, it is evaluated in the same way as in
Eq.~(\ref{eq:60}), by finding the {\it modified} complex trajectory
${\bf x}_\epsilon(t)$.\footnote{Since ${\bf x}_\epsilon(t)\in
  \mathbb{C}$, the functional $T_{int}$ should be continued into the
  complex domain. This is achieved by approximating $\theta(x)$ in
  Eq.~(\ref{eq:2}) by a smooth function $\theta_a(x)$; the width $a$
  of $\theta_a(x)$ is sent to zero in the final semiclassical
  expressions. Note that the value of $T_{int}$ is complex on complex
  paths.} By construction, ${\bf x}_\epsilon(t)$ extremizes the
modified action and satisfies the initial conditions (\ref{eq:18}).

The third favorable property of Eq.~(\ref{eq:28}) is an explicit
dependence of the integration limit on $\tau$. Let us
schematically explain how this feature simplifies evaluation of
the distribution $\rho(\tau)$. Recall that $\rho(\tau)$ is
proportional to the {\it derivative} of ${\cal P}_\tau$ with respect
to $\tau$, see Eq.~(\ref{eq:1}). This derivative is evaluated easily
if we represent ${\cal P}_\tau$  in the form ${\cal P}_\tau =
\int_0^{\tau} d\tau_+ \dots$,  where $\tau_+$ is an integration
variable, and the integrand does not depend on $\tau$. Then,
$\rho(\tau)$ is simply proportional to the integrand of ${\cal
  P}_\tau$ taken at $\tau_+=\tau$.

We derive the desired integral for ${\cal P}_\tau$ in Appendix
\ref{sec:modification}. We start from Eq.~(\ref{eq:28})
and perform all saddle--point integrations except for the integral
over $\tau'$ which is kept intact until the last moment. Then, we
substitute $\Psi(\tau,{\bf x}_f)$ into the probability
expression~(\ref{eq:55}) and integrate over ${\bf x}_f$. In this way
we obtain ${\cal P}_\tau$ in the integral form, where $\tau$
explicitly enters the integration limits. At this stage  the integrand
of ${\cal P}_\tau$ is a functional of the modified  
trajectory ${\bf x}_\epsilon(t)$ which is defined in the interval
$t\in [0;\,\tau]$. Using the modified equations of motion, we 
extend ${\bf x}_\epsilon(t)$ to $t\in [0;\, +\infty]$ and show that
motion at $t>\tau$ does not contribute into the integrand of ${\cal
  P}_\tau$. In this way we show that the integrand does not depend on
$\tau$. Thus, $\rho(\tau)$ is precisely the integrand of ${\cal
  P}_\tau$ at the end--point of integration; 
it is conveniently computed using the extended trajectory ${\bf
  x}_\epsilon(t)$, $t\in [0;\, +\infty]$. 

The result of the derivation sketched above (see
Appendix~\ref{sec:modification} for details) is summarized as  
follows. One starts by finding the modified trajectory ${\bf
  x}_\epsilon(t)$ which satisfies equations of motion following
from the modified action
\begin{equation}
\label{eq:72}
S_\epsilon[{\bf x}] = S[{\bf x}] + \frac{i\epsilon}{g^2} \,
(T_{int}[{\bf x}]-\tau)
\end{equation}
and old boundary conditions, Eqs.~(\ref{eq:18}), (\ref{eq:19}). 
Importantly, ${\bf x}_\epsilon(t)$ is extended beyond the interval
$t\in [0;\, \tau]$, and the final boundary conditions (\ref{eq:19})
are imposed at $t\to +\infty$.  Then, we determine the saddle--point
value of $\epsilon$. One can show that $\epsilon$ is real, so that the
modification term in Eq.~(\ref{eq:72}) is imaginary. This feature is
specific to the case of classically forbidden transitions. Besides,
$\epsilon$ is related to the duration $\tau$ by non--linear equation
\begin{equation}
\label{eq:78}
\mathrm{Re}\,T_{int}[{\bf x}_\epsilon] = \tau\;,
\end{equation}
which is similar to the respective constraint in the classically
allowed case. We remark in passing that $\epsilon=0$ corresponds to
the {\it original} complex trajectory ${\bf x}(t)$ which saturates the
probability ${\cal P}_\infty$, see Eq.~(\ref{eq:72}).

After finding the modified trajectory ${\bf x}_\epsilon(t)$ and
solving Eq.~(\ref{eq:78}) one computes the traversal--time
distribution,
\begin{equation}
\label{eq:75}
\rho(\tau) = {\cal N}
\sqrt{-\frac{d\epsilon}{d\tau}} \cdot A_\epsilon
\,\mathrm{e}^{-F_\epsilon/g^2}\;,
\end{equation}
where ${\cal N}$ is the normalization factor. Suppression exponent
$F_\epsilon$ and prefactor $A_\epsilon$ in Eq.~(\ref{eq:75}) are 
given by the same expressions as in Eq.~(\ref{eq:59}), but with the
modified action $S_\epsilon$ and modified trajectory ${\bf
  x}_\epsilon(t)$.

Note that the factor $A_\epsilon \cdot \exp(-F_\epsilon/g^2)$ entering
Eq.~(\ref{eq:75}) has meaning of ``naive'' transition probability
computed for the trajectory with action~(\ref{eq:72}).  The factor
$\sqrt{-d\epsilon/d\tau}$ is not so trivial, however. From the
technical viewpoint it arises due to fixation of $T_{int}$ in the path
integral~(\ref{eq:60}). 

In what follows we exploit the relation which is derived in
Appendix~\ref{sec:modification},
\begin{equation}
\label{eq:79}
2\epsilon = - dF_\epsilon / d\tau\;.
\end{equation}
Equation~(\ref{eq:79}) shows that the values of $\epsilon$ and $\tau$
are related by Legendre transformation. This is what one expects,
since $\epsilon$ is interpreted as a Lagrange multiplier conjugate to
$\tau$.

It is worth noting that the method of ``Lagrange multipliers''
introduced above resembles the approach of
Refs.~\cite{Sokolovski:2008}. Note, however, one important difference:
we define traversal--time distribution $\rho(\tau)$ by
Eq.~(\ref{eq:1}) and then introduce the functional $T_{int}[{\bf x}]$  as
a convenient technical tool for the semiclassical calculation of this
distribution. The authors of Refs.~\cite{Sokolovski:2008} go the other
way around. They start with the functional similar to Eq.~(\ref{eq:2})
and then {\it define} traversal time using path integral technique. 

Let us mention one important restriction of the semiclassical method
of this Section. While deriving Eq.~(\ref{eq:75}) we assumed that only
one complex trajectory gives substantial contribution into 
$\rho(\tau)$. This assumption is certainly correct for
one--dimensional conservative systems; explicit calculations show that
it also holds\footnote{In a wide interval $\tau\in \Delta \tau$ centered
  around  $\langle \tau \rangle$. Typically, the width of this interval
  is of order $40$ and becomes somewhat smaller in the vicinity of the
  critical point $p_0 \approx p_{c,1}$. In what follows we work at
  $\tau \in \Delta \tau$ and use Eq.~(\ref{eq:75}).} in the
two--dimensional model (\ref{eq:58}). In other multidimensional  
systems~\cite{Takahashi:2006} there might exist several (or
even an infinite number of) trajectories giving comparable
contributions into the traversal--time distribution. In this case 
Eq.~(\ref{eq:75}) should be modified: contributions of different
trajectories and respective interference terms should be accounted
for. 

We finish this Section by considering the standard tunneling mechanism
where the original complex trajectory ${\bf x}(t)$ is stable. We are
going to show that in this case $\rho(\tau)$ is
symmetric and sharply peaked around the central value $\tau = \langle \tau
\rangle$. From the Legendre transformation (\ref{eq:79}) one
sees that the exponent $F_\epsilon$ in
Eq.~(\ref{eq:75}) has minimum at $\epsilon(\tau) = 0$. This
point corresponds to the {\it original} trajectory ${\bf x}(t)$. Since 
${\bf x}(t)$ is stable, the respective value of $\tau= \mathrm{Re}\,
T_{int}[{\bf x}]$ is finite. One expands
the exponent around the point $\epsilon(\tau)=0$ and obtains the  
Gaussian distribution, 
\begin{equation}
\label{eq:80}
\rho_s(\tau) = \frac{1}{\sigma_\tau\sqrt{2\pi}}
\cdot\mathrm{e}^{-(\tau - \langle \tau \rangle)^2/2\sigma_\tau^2}\;,
\end{equation}
where we restored the normalization factor and added subscript $s$
(for ``stable''). The central value $\langle \tau \rangle$ and dispersion
$\sigma_\tau^2$ of this distribution are given by the expressions
\begin{equation}
\label{eq:81}
\langle \tau \rangle = \mathrm{Re}\,T_{int}[{\bf x}] \sim O(g^0)\;,
\qquad \qquad \sigma_\tau^2 =
-\frac{g^2}{2}\frac{d\tau}{d\epsilon}\Bigg|_{\epsilon=0} \sim O(g^2)
\end{equation}
which come from the Taylor series expansion of the semiclassical
exponent and Eqs.~(\ref{eq:78}), (\ref{eq:79}). We remark that
Eq.~(\ref{eq:80}) is applicable at $|\tau - \langle \tau \rangle| \sim
g$ where $\rho(\tau) \sim 1$. Outside of this region the
value of $\rho(\tau)$ is negligibly small. Note also that the average
duration 
$\langle \tau \rangle$ in Eqs.~(\ref{eq:81}) is equal to the time
spent by the {\it original} complex trajectory at $x<x_2$. 

The above considerations are not valid if the original
trajectory ${\bf x}(t)$ is unstable. In that case the point
$\epsilon= 0$ corresponds to $\tau = +\infty$ and the 
semiclassical exponent in Eq.~(\ref{eq:75}) remains almost constant as
$\tau\to +\infty$.  Thus, $\rho(\tau)$ is essentially asymmetric: at
large $\tau$ it falls off slowly, only due to prefactor. In the next
Section we will show that the corresponding values  of $\langle \tau
\rangle$ and $\sigma_\tau^2$ are large.

\section{Results}
\label{sec:results}
In this Section we compute distributions $\rho(\tau)$ for three
quantum processes related to unstable semiclassical dynamics.

\subsection{One dimension}
\label{sec:one-dimension-1}
We begin by considering one--dimensional activation transitions
introduced in Sec.~\ref{sec:sphal-driv-tunn}. For definiteness we
suppose that the potential $U(x)$ reaches its maximum at $x=0$. 
We denote by $\omega_-^2 = -U''(0)$ the curvature of the potential at
the maximum. Note that $\omega_-$ stays finite as $g^2\to 0$. We
will also use the exemplary potential
\begin{equation}
\label{eq:50}
U(x) = \frac{1}{g^2\, \mathrm{ch}^2 (gx)}
\end{equation}
for checking and illustrating the semiclassical technique.

Let us show that the semiclassical dynamics is unstable in the case of
activation transitions. We use the logic of
Sec.~\ref{sec:sphal-driv-tunn} and decompose the
initial state $\Psi_i$ in the basis of plane waves with fixed momenta
$p$. Denoting by ${\cal P}_p$ the transmission coefficients of
individual waves, we write,
\begin{equation}
\label{eq:51}
{\cal P}_\infty = \int_0^{\infty} dp \;{\cal P}_p \cdot |\Psi_i(p)|^2
= \int_0^{\infty} dp \;{\cal P}_p
\cdot\frac{1}{\sigma_p\sqrt{2\pi}}
\;\mathrm{e}^{-{(p-p_0)^2}/{2\sigma_p^2}}\;,
\end{equation}
where interference terms are absent due to conservation of energy
$E_p = p^2/2$. Recall that the value of $p_0$ is smaller than
$p_{c,2}\equiv \sqrt{2V_0}$. On the other hand, the
integration momentum $p$ in Eq.~(\ref{eq:51}) can be higher or lower
than $\sqrt{2V_0}$ representing the ``over--barrier'' or
``under--barrier'' 
modes, respectively. We consider these cases separately. Transmissions
at $p>\sqrt{2V_0}$ are described by classical over--barrier
trajectories. Thus, ${\cal P}_p = 1$ and 
the integrand in Eq.~(\ref{eq:51}) decreases with $p$ at
$p>\sqrt{2V_0}$ due to the in--state contribution.

At $p<\sqrt{2V_0}$ the situation is different, since the  coefficients
${\cal P}_p$ are exponentially suppressed. The semiclassical
expression for these coefficients is well--known, ${\cal P}_p =
\mathrm{exp}[-S(E_p)]$ where $S(E_p)$ is Euclidean action computed
for one period of Euclidean oscillations in the upside--down 
potential. The semiclassical trajectory $x(t)$ performing oscillations 
has energy $E_p=p^2/2$. Let us visualize how $x(t)$ moves between
the in-- and out-- asymptotic regions. In the in--region the
trajectory evolves in real time until it stops at the left turning
point of the potential. The subsequent semiclassical evolution
proceeds in Euclidean time where $x(t)$ performs precisely one
half--period oscillation in the upside--down potential. At the end of
oscillation the trajectory arrives at the right turning point. After
that $x(t)$ moves in real time, again, and finally ends up in the
out--region. Only the Euclidean part of $x(t)$ contributes into ${\cal
  P}_p$. 

One sees that for under--barrier modes both factors in the integrand
of Eq.~(\ref{eq:51}) are exponentially sensitive to the momentum
$p$. To understand the behavior of the integrand, we consider 
its logarithmic derivative,
\begin{equation}
\label{eq:3}
\frac{d}{dp} \mathrm{ln}\left[ {\cal P}_p |\Psi_i(p)|^2 \right]
 = p\,T(E_p) -  \frac{p-p_0}{\sigma_p^2}\;,
\end{equation}
where $T(E) = -dS/dE$ is the period of Euclidean oscillations. 
Now, we take the limit $p\to \sqrt{2V_0}$ in 
Eq.~(\ref{eq:3}). In this limit  $E_p$ approaches the height
of the potential barrier, and Euclidean oscillations get
restrained to a small region near the barrier top. Thus, $T\to
2\pi/\omega_-$. Using this property, one finds that the derivative
(\ref{eq:3}) is {\it positive} at $p=\sqrt{2V_0}$ if $p_0 > p_{c,1}$,
where 
\begin{equation}
\label{eq:4}
p_{c,1} =\sqrt{2V_0} \,(1- 2\pi\sigma_p^2/\omega_-) 
\end{equation}
is the critical momentum. Note that in the limit $E_p\to V_0$ the
semiclassical trajectory $x(t)$ becomes {\it unstable}: it starts
in the in--region and ends up on top of the potential barrier as $t\to 
+\infty$.

Now, one recalls that the integrand in
Eq.~(\ref{eq:51}) decreases with $p$ at $p>\sqrt{2V_0}$. Thus, 
it has sharp maximum at $p=\sqrt{2V_0}$ whenever
$p_0$ lies between $p_{c,1}$ and $p_{c,2}\equiv\sqrt{2V_0}$. Moreover,
this maximum is global\footnote{The case of ``pathological''
  potentials with non--monotonic $T(E)$ should be considered
  separately.} if $T(E)$ decreases with energy, see Eq.~(\ref{eq:3}).
Thus, at $p_{c,1} < p_0 < p_{c,2}$ the integral for the
transmission probability ${\cal P}_\infty$ is saturated at $E_p =
V_0$, i.e. precisely in the vicinity of unstable semiclassical
trajectory.

The above considerations deserve two remarks. First, one can show
that the semiclassical trajectory saturating the probability
(\ref{eq:51}) satisfies the boundary conditions of the previous
Section, Eqs.~(\ref{eq:18}), (\ref{eq:19}). This is natural, since
Eqs.~(\ref{eq:18}), (\ref{eq:19})  are obtained from the requirement 
that the probability is extremal, see
Appendix \ref{sec:standard-method}. In Appendix~\ref{sec:1d} we
explicitly solve the semiclassical equations in the
model~(\ref{eq:50}) and find the trajectory $x(t)$. It is unstable if
$p_{c,1} < p_0 < p_{c,2}$.

Second, the interval of unstable motions is semiclassically
large, $p_{c,2}-p_{c,1} \sim 1/g$, and shrinks to a point as
$\sigma_p^2\to 0$. This means that unstable semiclassical dynamics is
essential in the case $g^2\ll 1$, $\sigma_p^2\sim 1$ which we consider.

Let us compute $\rho(\tau)$ in the case of unstable
semiclassical trajectories $x(t)$. We have learned that such
trajectories do not interpolate between the in-- and out--
regions of the process, but start in the in--region at $t=0$ and
approach exponentially the barrier top $x=0$ as $t\to +\infty$. At
large times one writes $x(t) = c_- \mathrm{e}^{-\omega_-t}$, where
$c_-$ is a complex constant. Following the method of the previous
Section, we introduce modification, 
Eq.~(\ref{eq:72}). We will restrict attention to the case of small
$\epsilon$ which will be important for calculating $\rho(\tau)$. One
notes that the modified trajectory $x_\epsilon(t)$ moves in the {\it
  complex} potential 
\begin{equation}
\label{eq:22}
U_\epsilon(x) = U(x)-\frac{i\epsilon}{g^2}\,\theta(x_2-x)\;.
\end{equation}
It cannot stop at the barrier top due to conservation of energy,
which is real by Eqs.~(\ref{eq:19}). The evolution described by 
$x_\epsilon(t)$  proceeds in three stages: the trajectory arrives into
the region $x\approx 0$, spends some time there and then departs for
the out--region. The modification term can be neglected during the
first and last stages of evolution, but destroys unstable motion
near the  barrier top. Indeed, at $x_\epsilon\approx 0$ one writes,
\begin{equation}
\label{eq:11}
x_\epsilon(t) = c_{-}\, \mathrm{e}^{-\omega_- t} + \epsilon c_+\,
\mathrm{e}^{+\omega_- t}\;, 
\end{equation}
where $\epsilon$ in the second term reflects the fact that
the original trajectory  
is recovered in the limit $\epsilon\to 0$. Since $\epsilon\ll 1$, the
time of motion near the barrier top is large, and one computes 
$T_{int}$, Eq.~(\ref{eq:78}), using the evolution
(\ref{eq:11}), 
\begin{equation}
\label{eq:12}
\tau = -\frac{1}{\omega_-}\mathrm{ln}(\epsilon/\epsilon_0) +
O(\epsilon)\;,
\end{equation}
where $\epsilon_0\sim O(g^0)$ is some constant. In practice one
finds the value of this constant explicitly, by obtaining the
trajectory at all three stages of the modified evolution
and calculating $T_{int}[x_\epsilon]$.  Given $\epsilon(\tau)$, one
computes the suppression exponent from Eq.~(\ref{eq:79}),
$$
F_\epsilon = F_\infty + 2\int_\tau^{+\infty} d\tau' \,\epsilon(\tau')
= F_\infty + \frac{2\epsilon_0}{\omega_-} \mathrm{e}^{-\omega_-
  \tau}\;, 
$$
where $F_\infty$ is independent of $\tau$.

Now, we estimate the modified prefactor $A_\epsilon$ at $\epsilon\ll
1$. The procedure for evaluating $A_\epsilon$ is described in
Appendix~\ref{sec:1d}; we proceed by following this procedure in the
case of small $\epsilon$. One finds the perturbation
$\psi_\epsilon(t)$ which satisfies the linearized equations of motion
in  the background of the modified trajectory
$x_\epsilon(t)$. Boundary  conditions for $\psi_\epsilon(t)$ are
imposed in the out--region. Evolving the perturbation back in time,
one  notes that $\psi_\epsilon(t)$ grows exponentially during the
second stage of the modified evolution, Eq.~(\ref{eq:11}). Thus, it
becomes exponentially large at $t=0$, $\psi_\epsilon(0) \sim
\mathrm{exp}(\omega_- \tau)$. At $\epsilon\ll 1$ the prefactor is
estimated\footnote{We 
  disregarded irrelevant numerical factors in Eq.~(\ref{eq:21}) and
  perturbation $\chi_\epsilon(t)$ which does not grow exponentially.} as
$A_\epsilon \sim [\psi_\epsilon(0)]^{-1/2}$, see Eq.~(\ref{eq:21}) of
Appendix~\ref{sec:1d}. One finds $A_\epsilon \sim
\mathrm{exp}(-\omega_- \tau/2)$.

Substituting $\epsilon(\tau)$, $F_\epsilon$ and $A_\epsilon$ into
Eq.~(\ref{eq:75}), we obtain the distribution
\begin{equation}
\label{eq:8}
\rho_u(\tau) = \frac{2\epsilon_0}{ g^2}\, \mathrm{exp}\left\{ -\omega_-
  \tau  - \frac{2\epsilon_0}{g^2\omega_-}\,\mathrm{e}^{-\omega_-
    \tau}\right\}\;,
\end{equation}
which has the form of Gumbel distribution of
type~I~\cite{Gumbel:1958}. We will argue below that Eq.~(\ref{eq:8})
is valid in the region $\tau \gg |\log 
\epsilon_0|/\omega_-$ where $\rho_u(\tau)$ is not exponentially
small. By using the subscript $u$ in the notation $\rho_u(\tau)$ we
stress that Eq.~(\ref{eq:8}) is valid in the case of unstable
semiclassical trajectories.

Several remarks  are in order. First, $\rho_u(\tau)$ is drastically
different from the respective Gaussian distribution (\ref{eq:80})
which is valid in
the case of stable semiclassical trajectories. Indeed,
function~(\ref{eq:8}) is essentially asymmetric. Besides, it
corresponds to large mean time of transmission and large
time dispersion, 
\begin{equation}
\label{eq:10}
\langle \tau \rangle = \frac{1}{\omega_-} \mathrm{ln}\left[
    \frac{2\epsilon_0}{g^2\omega_-}\right] + \frac{\gamma}{\omega_-}
  \;, \qquad \qquad \sigma_\tau^2 = \frac{\pi^2}{6\omega_-^2}\;,
\end{equation}
where $\gamma$ is the Euler constant. The quantities (\ref{eq:10})
depend on 
the semiclassical parameter as $\langle \tau \rangle \sim |\log g^2|$,
$\sigma_\tau^2 \sim O(g^0)$, in contrast to the respective
scalings $\langle \tau \rangle \sim O(g^0)$, $\sigma_\tau^2 \sim g^2$
in the case of stable trajectories, cf. Eqs.~(\ref{eq:81}). 

Second, the form of the distribution~(\ref{eq:8})  is
{\it universal}. Indeed, in deriving $\rho_u(\tau)$ we did not use
any information about the potential $U(x)$ or initial wave packet
$\Psi_i(x)$. All 
this information is encoded in two parameters of the distribution,
$\epsilon_0$ and $\omega_-$. Moreover, one represents Eq.~(\ref{eq:8}) 
in the form
$$
\rho_u(\tau) = \omega_- \exp\left\{-\omega_-(\tau - \langle \tau
  \rangle) - \gamma - \mathrm{e}^{-\omega_-(\tau - \langle \tau \rangle)
    - \gamma}\right\}\;,
$$
where the first of Eqs.~(\ref{eq:10}) was used.
All parameters in this formula disappear after
appropriate shifts and rescalings of $\tau$. 

Third, during the derivation of Eq.~(\ref{eq:8}) we assumed that
$\epsilon\ll 1$. This is legitimate at $\tau \gg |\log
\epsilon_0|/\omega_-$, see Eq.~(\ref{eq:12}). In particular, 
the region $\rho_u(\tau) \sim  1$ corresponds to $\tau \sim \langle
\tau \rangle$ which leads to $\epsilon \sim g^2$,
cf. Eqs.~(\ref{eq:10}) and (\ref{eq:12}).

\begin{figure}[t]
\centerline{\includegraphics[width=0.5\textwidth]{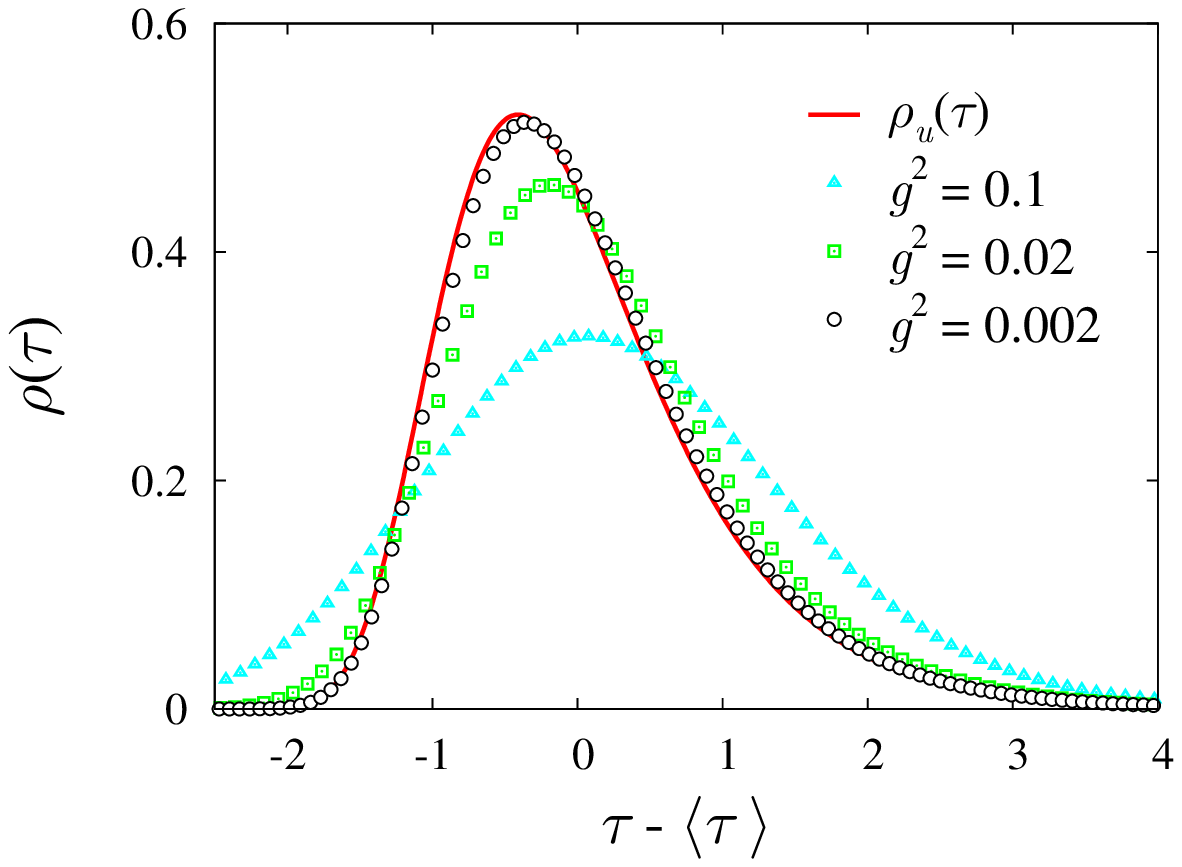} 
\includegraphics[width=0.5\textwidth]{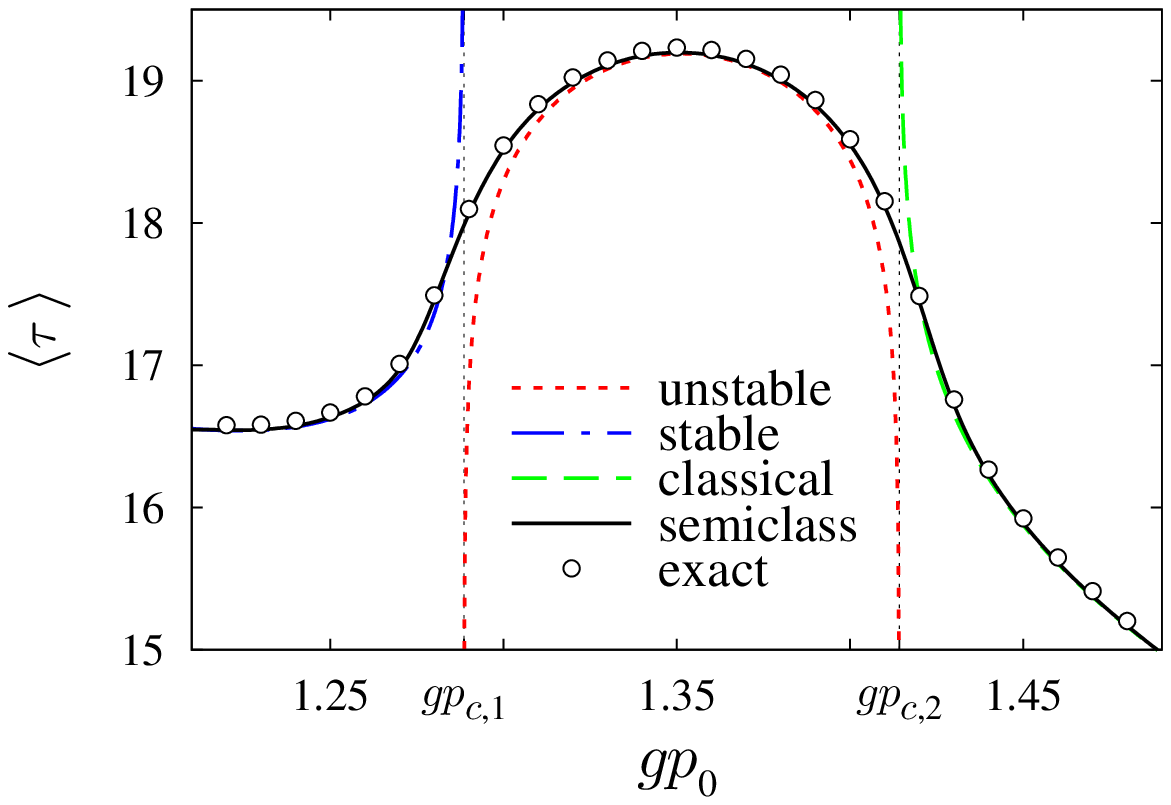}}
\centerline{\hspace{1cm}(a) \hspace{7.5cm} (b)}
\caption{\label{fig:6}(Color online)  (a) One--dimensional traversal--time
  distributions in the case of unstable semiclassical dynamics. Points
  and solid line represent exact quantum mechanical results and
  Eq.~(\ref{eq:8}), respectively. (b) Average time of passing in the model 
  (\ref{eq:50});  $g^2 = 0.002$, $\sigma_p^2 = 0.02$, $gx_2 = -gx_1 =
  10$.}
\end{figure}
We derived expression~(\ref{eq:8}) using the properties of the
modified semiclassical dynamics. This expression is illustrated in
Appendix~\ref{sec:1d} where we compute 
$\rho(\tau)$ for the model 
(\ref{eq:50}). In the region $p_{c,1} < p_0 < p_{c,2}$ 
the result of Appendix~\ref{sec:1d} coincides with Eq.~(\ref{eq:8}),
where\footnote{Exponential dependence of $\epsilon_0$ on $x_2-x_1$
has simple physical meaning. Namely, Eqs.~(\ref{eq:10}), (\ref{eq:13})
imply that $\langle \tau \rangle = g(x_2-x_1)/\sqrt{2} + 
\dots$, in accordance with the fact that the modes saturating the
total transmission probability move with momentum $p\approx\sqrt{2V_0} =
\sqrt{2}/g$.} 
\begin{equation}
\label{eq:13}
\epsilon_0 = \sin \left[\frac{\sqrt{2}-gp_0}{2\sigma_p^2}\right]
\cdot \mathrm{e}^{g(x_2-x_1)}\;.
\end{equation}
Note that $\epsilon_0$ vanishes at $p=p_{c,1}$ and
$p= p_{c,2}\equiv \sqrt{2}/g$, see Eq.~(\ref{eq:4}). Function
(\ref{eq:8}) is depicted in 
Fig.~\ref{fig:6}a, solid line\footnote{We plot the distributions as  
  functions of $\tau - \langle \tau \rangle$; in this case all
  parameters except for $\omega_-$ vanish from
  Eq.~(\ref{eq:8}). Besides, we use $\omega_- = \sqrt{2}$ which is
  valid for the potential (\ref{eq:50}).}.

As a check of the modified semiclassical method, we compare  
 $\rho_u(\tau)$ with exact quantum mechanical results
(points in Fig.~\ref{fig:6}a). The latter results are obtained by
propagating numerically\footnote{This can be achieved by Fourier
  methods~\cite{Rebbi} or by using the basis of energy
  eigenstates in the model~(\ref{eq:50}). We checked that both methods
  produce equivalent results.} wave packets in the
model~(\ref{eq:50}) and using Eqs.~(\ref{eq:55}), (\ref{eq:1}). One
sees that the exact graphs are almost symmetric at large $g^2$
but change their form and approach $\rho_u(\tau)$ as $g^2\to 0$. 

Our semiclassical expressions for the average time of passing are
summarized in Fig.~\ref{fig:6}b which displays dependence of $\langle
\tau \rangle$ on $p_0$ in the model~(\ref{eq:50}). The interval of unstable
semiclassical motions $p_{c,1} < p_0 < p_{c,2}$ is delimited on the
graph by the
vertical dotted lines. The regions to the left and to the
right of this interval correspond to stable tunneling trajectories and
classical over--barrier solutions, respectively. In accordance with
the above discussion, we compute $\langle \tau \rangle$ by
Eq.~(\ref{eq:10}) in the central part of the plot and
by Eq.~(\ref{eq:81}) in the rest of it. The  corresponding 
dependencies are marked in Fig.~\ref{fig:6}b as ``unstable'', ``stable'' and
``classical''. All three graphs exhibit unphysical behavior in the
vicinities of critical momenta $p_{c,1}$, $p_{c,2}$,  where the
``stable'' and  ``classical'' curves grow to infinity, and 
``unstable'' graph sharply drops down. This behavior is related to
``phase transition'' 
which transforms stable trajectories into unstable and Gaussian
distribution (\ref{eq:80}) into Eq.~(\ref{eq:8}). Near the critical
points $\langle \tau \rangle$ is computed with the full semiclassical
distribution (\ref{eq:75}). This calculation is performed in
Appendix~\ref{sec:1d}. In Fig.~\ref{fig:6}b we show the result by the
solid line which smoothly interpolates between the ``stable'',
``unstable'' and ``classical'' graphs and coincides with the
average time of passing extracted from the exact quantum mechanical
calculations (points).

\subsection{Two dimensions}
\label{sec:two-dimensions-1}
We proceed by considering the case of sphaleron--driven transitions 
in the two--dimensional model of Sec.~\ref{sec:sphal-driv-tunn}. At
the level of complex trajectories these transitions look similar to
one--dimensional activation processes of the previous
Section. However,  there is one important difference: the origin of
semiclassical instabilities in the sphaleron--driven case is related
to non--linear interaction between  the degrees of freedom, rather
than to the momentum dispersion in the initial state. Recall that the
new mechanism is relevant in the region $p_{c,1} < p < p_{c,2}$, where
$p_{c,1}$ and $p_{c,2}$ are critical momenta defined in
Sec.~\ref{sec:sphal-driv-tunn}. 

In the semiclassical calculations of this Section we use numerical
methods of Refs.~\cite{Bonini:1999kj,epsilon}. Namely, we
smoothen the step--function in the modified potential,
Eq.~(\ref{eq:22}),
$$
\theta(x) \to \theta_a(x) = \frac{1}{1+\mathrm{e}^{-x/a}}\;,
$$
where the width $a$ is supposed to be small\footnote{Practical
  calculations show that $\rho(\tau)$ is almost independent of $a$: at
  $a=0.1$ the values of $\rho$ stabilize at the
  level of accuracy $10^{-5}$. In what follows we use $a=0.1$.}. 
After that we introduce non--uniform lattice $t \in \{t_k |\, k=1\dots
N_k\}$. The trajectory ${\bf x}_\epsilon(t)$ is found from the
modified equations 
of motion and boundary conditions (\ref{eq:18}), (\ref{eq:19}) in the
discrete case.  Then, the values of $F_\epsilon$,
$A_\epsilon$ and $\rho(\tau)$ are computed by Eqs.~(\ref{eq:65}),
(\ref{eq:67}) of Appendix~\ref{sec:standard-method} and
Eq.~(\ref{eq:75}), respectively. 

We start with the simplified case when the total energy $E$ of
the original sphaleron--driven trajectory ${\bf x}(t)$ is close to the
height of 
the barrier, $E-V_0 \ll V_0$. The sphalerons with these energies
describe small linear oscillations around the saddle point $x=y=0$ of
the potential; for the modified semiclassical motion in the vicinity
of the saddle point one writes, 
\begin{equation}
\label{eq:24}
{\bf x}_\epsilon(t) = {\bf x_-} \left(c_- \mathrm{e}^{-\omega_- t} +
\epsilon c_+ \mathrm{e}^{\omega_- t}\right) + {\bf x_+} c
\cos(\omega_+ t + \varphi) \;,
\end{equation}
cf. Eq.~(\ref{eq:11}). Vectors ${\bf x}_+$ and ${\bf x}_-$ in this
expression run along ``stable'' and ``unstable'' directions of the saddle
point, while $\omega_\pm$ represent respective ``frequencies.''
Using the same arguments
as in the previous Section, one shows that the 
distribution $\rho(\tau)$ is given by
Eq.~(\ref{eq:8}) if Eq.~(\ref{eq:24}) is justified (i.e. at $E-V_0 \ll
V_0$).

\begin{figure}[htb]
\centerline{\includegraphics[width=0.5\textwidth]{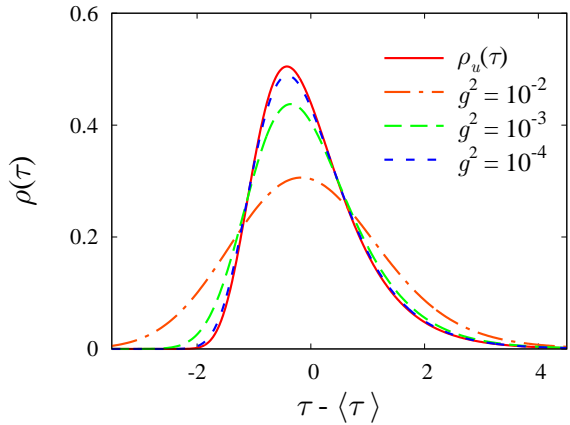}}
\caption{\label{fig:3}(Color online) Comparison between
  one--dimensional semiclassical formula (\ref{eq:8}) (solid line)
  and semiclassical traversal--time distributions in the case of
  sphaleron--driven tunneling (dashed lines). The parameters of the
  graphs are $gp_0 = 1.038$, $g^2E_y = 0.45$; $\sigma_p^2 = 0.005$,
  $gx_2 = -gx_1 = 20$.}
\end{figure}
In Fig.~\ref{fig:3} we compare
  one--dimensional formula\footnote{We find $\rho_u(\tau)$ using
  the ``negative'' eigencurvature $\omega_-^2 = 1-\omega^2/2 + \sqrt{1
  + \omega^4/4}$ of the potential~(\ref{eq:58}) at the saddle point.},
  Eq.~(\ref{eq:8}), with two--dimensional 
  semiclassical distributions~(\ref{eq:75}) at different values of
  $g^2$. Note that all distributions in Fig.~\ref{fig:3} are 
  semiclassical; nevertheless, their forms depend on $g^2$. One
  explains this as follows. We selected parameters of
  Fig.~\ref{fig:3} in such a way that the energy $E$ of 
  the original trajectory ($\epsilon=0$) is almost equal to the
  barrier height, $(E - V_0)/V_0 \sim 10^{-5}$. On the other hand, the
  energies of ${\bf 
    x}_\epsilon(t)$ remain close to $V_0$ at
  small $\epsilon$ (large $\tau$) and depart from it at $\epsilon \sim
  1$ (smaller $\tau$). Thus, linear approximation (\ref{eq:24}) breaks
  down if the value of $\tau$ is not sufficiently large. Due to this
  fact the semiclassical distributions in Fig.~\ref{fig:3} coincide
  with $\rho_u(\tau)$ only at small $g^2$. Indeed the central parts
  of the distributions correspond to $\tau \approx \langle \tau
  \rangle \sim |\log g^2|$, see Eq.~(\ref{eq:10}); at large $g^2$ the
  distributions shift\footnote{This is not seen in Fig.~\ref{fig:3} 
  where the distributions are functions of $\tau - \langle \tau
  \rangle$.} to smaller $\tau$, where Eq.~(\ref{eq:24}) gets
violated.

\begin{figure}
\centerline{\includegraphics[width=0.5\textwidth]{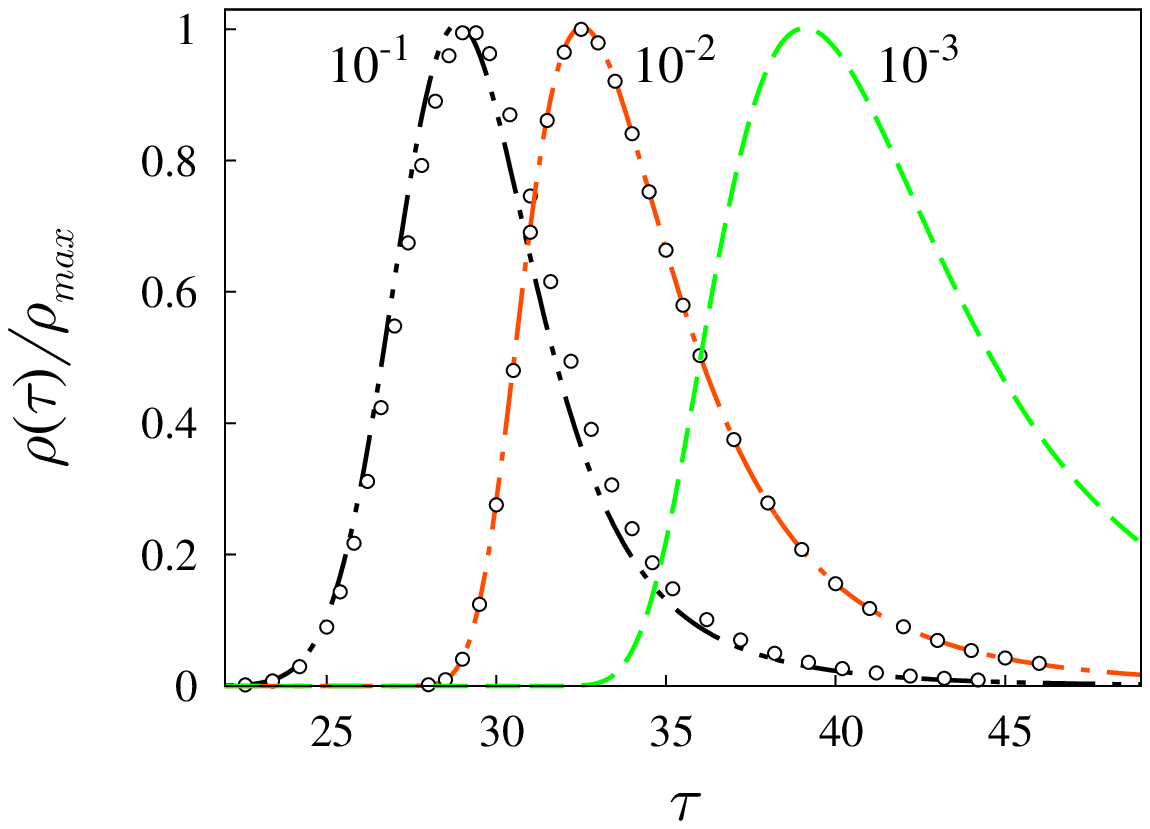}
\includegraphics[width=0.5\textwidth]{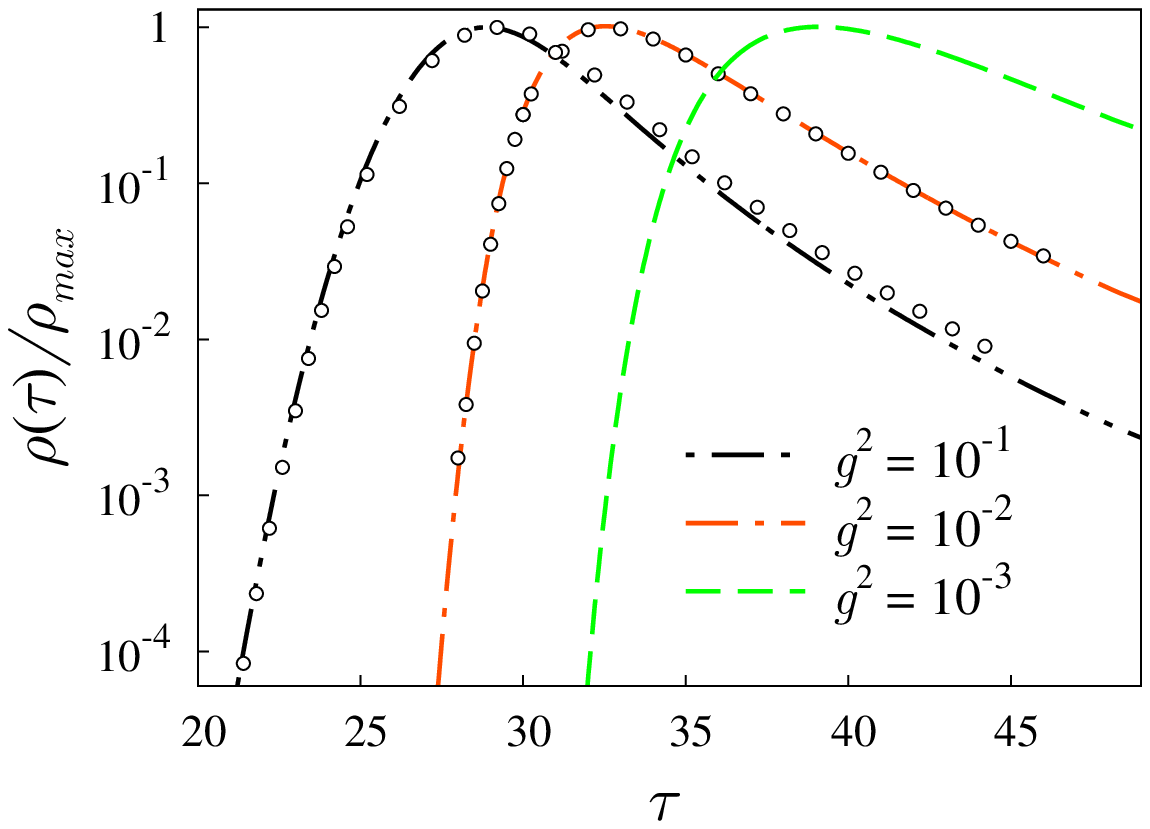}}
\centerline{\hspace{1cm} (a) \hspace{7.5cm} (b)}
\caption{\label{fig:5}(Color online) (a) Semiclassical (lines) and
  exact quantum--mechanical (points) traversal--time distributions in
  the model (\ref{eq:58}). The distributions are plotted in units of
  their maximal values $\rho_{max}$. Numbers near the graphs give 
  values of $g^2$; the other parameters are $gp_0 = 1.52$,
  $g^2E_y = 0.05$; $\sigma_p^2 = 0.005$, $gx_2 = -gx_1 = 20$. (b) The
  same as (a) but in logarithmic scale.}
\end{figure}
In general case when the energy $E$ of the original unstable
trajectory ${\bf x}(t)$ is substantially higher than $V_0$ the
distribution $\rho(\tau)$ is not given by Eq.~(\ref{eq:8}), regardless 
of whether $g^2$ is small or not. However, many qualitative features of
$\rho_u(\tau)$ are valid in the higher--energy case as well. Consider
e.g. Fig.~\ref{fig:5}a which corresponds to $(E-V_0)/V_0 \approx 0.2$. 
The distributions in this figure are highly asymmetric; they have long
exponential tails at large $\tau$ and steep front--ends. Exponential
behavior of $\rho(\tau)$ at large $\tau$ is illustrated in
Fig.~\ref{fig:5}b where Fig.~\ref{fig:5}a is replotted in logarithmic
scale. One writes,
\begin{equation}
\label{eq:36}
\rho(\tau) \propto \exp(-\mbox{const}\cdot \tau)\;,
\end{equation}
since the graphs in Fig.~\ref{fig:5}b are almost linear at $\tau
\gg \langle \tau \rangle$.

In Figs.~\ref{fig:5} we compare the semiclassical (lines) and exact 
quantum--mechanical (points) results for the traversal--time 
distributions; one observes agreement. The exact results are
obtained by evolving numerically wave packets in full quantum system
and implementing Eqs.~(\ref{eq:55}), (\ref{eq:1}); we describe the
respective numerical method in
Appendix~\ref{sec:exact-quant-mech}. 

\begin{figure}
\centerline{\includegraphics[width=0.5\textwidth]{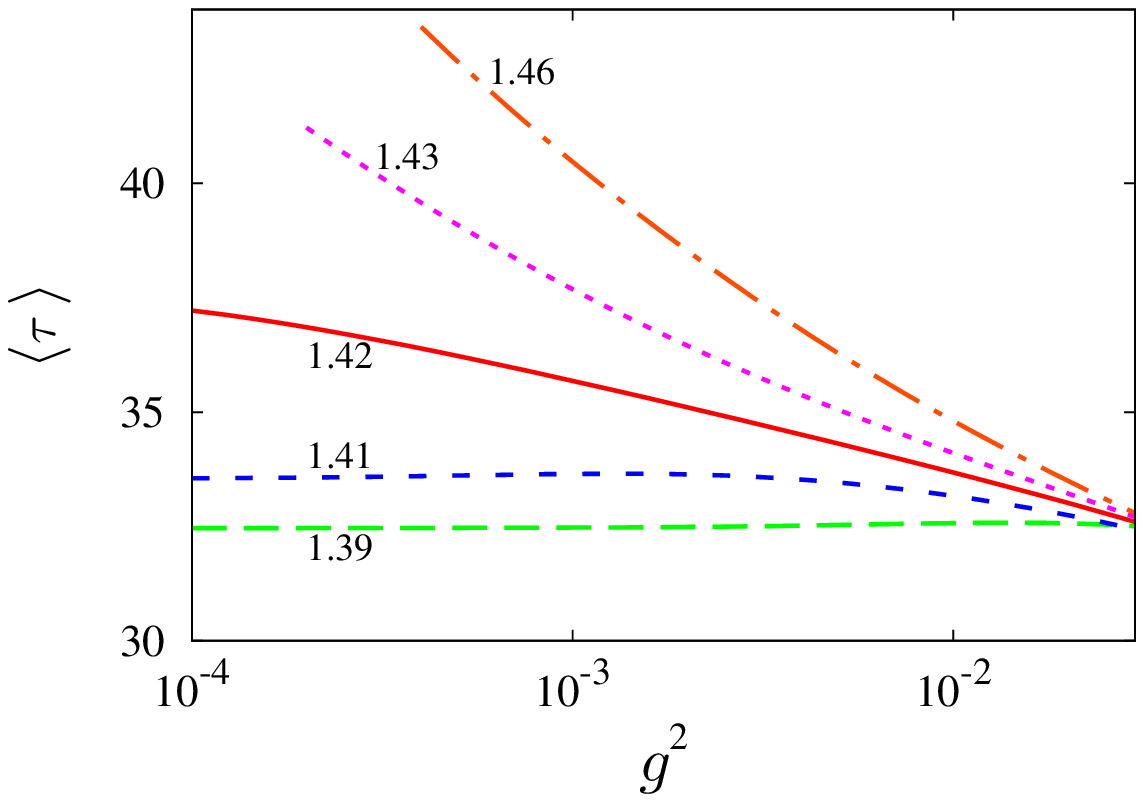}
\includegraphics[width=0.5\textwidth]{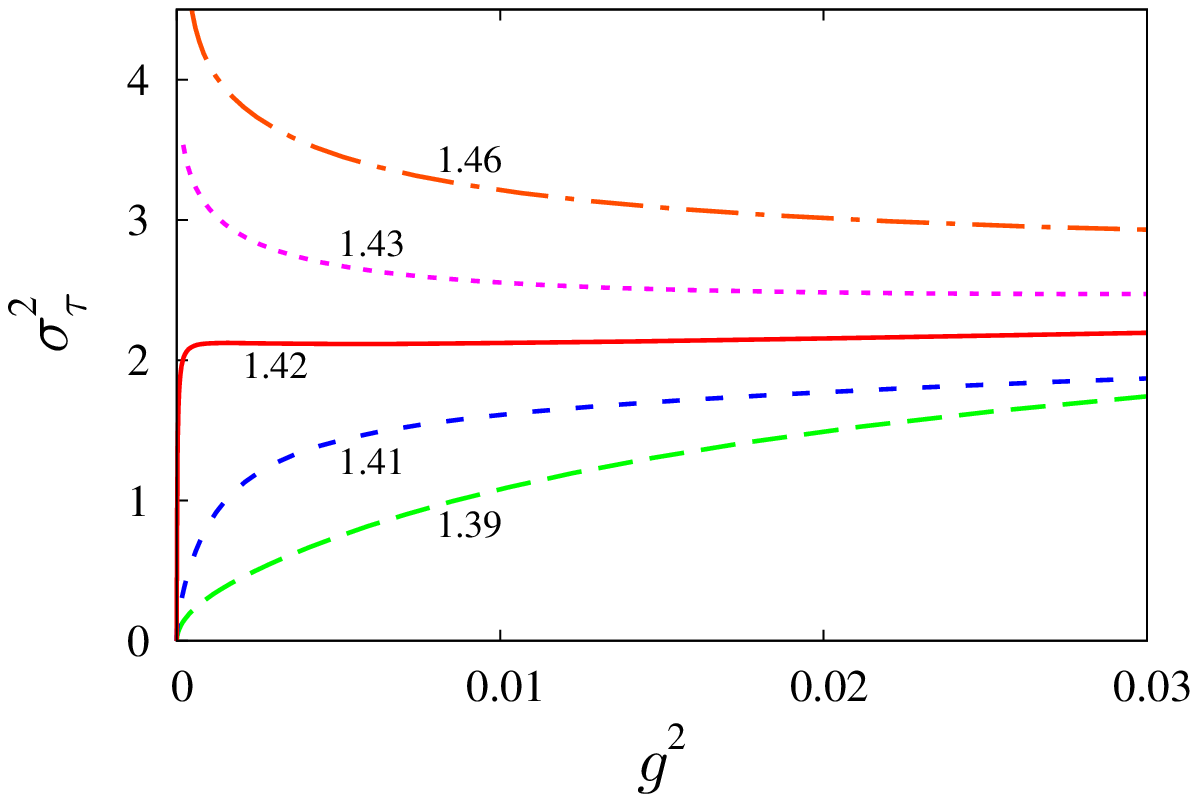}}
\centerline{\hspace{1cm} (a) \hspace{7.5cm} (b)}
\caption{\label{fig:10}(Color online) Mean times of passing (a) and
  time dispersions (b) as functions of 
  $g^2$ in the model (\ref{eq:58}). Numbers near the graphs give 
  values of $g p_0$. The other parameters are the same 
  as in Figs.~\ref{fig:5}. Note that $g p_{c,1} \approx 1.421$; graphs
  with $p_0 > p_{c,1}$ correspond to the case of sphaleron--driven
  tunneling.}
\end{figure}
The above properties of $\rho(\tau)$ naturally lead to large mean
times of transmission $\langle \tau \rangle$ and large time dispersions
$\sigma_\tau^2$ in the case of sphaleron--driven tunneling. In
Figs.~\ref{fig:10} we plot $\langle \tau \rangle$ and $\sigma_\tau^2$
as functions of $g^2$ at different values of $p_0$. One observes
distinct behavior in the cases $p_0 < p_{c,1}$ and $p_0 > p_{c,1}$
corresponding to stable and unstable semiclassical dynamics,
respectively. At small $p_0$ the graphs in Figs.~\ref{fig:10} agree
with Eqs.~(\ref{eq:81}): $\langle \tau \rangle$ 
remains constant, and $\sigma_p^2$ vanishes in the limit $g^2\to
0$. In the unstable case $p_0 > p_{c,1}$ one sees that $\langle \tau
\rangle \sim |\log g^2|$ while $\sigma_\tau^2$ grows as $g^2 \to 0$.
Using these dependencies, one can discriminate between the standard
and sphaleron--driven mechanisms of tunneling. Note that analogy
with one--dimensional activation processes suggests logarithmic growth 
of $\langle \tau \rangle$ but {\it constant} $\sigma_\tau^2$, see
Eqs.~(\ref{eq:10}). Apparently, the  unexpected behavior 
of $\sigma_\tau^2$ in Fig.~\ref{fig:10}b is related to features of
non--linear multidimensional dynamics which deserve separate study.

\begin{figure}
\centerline{\includegraphics[width=0.5\textwidth]{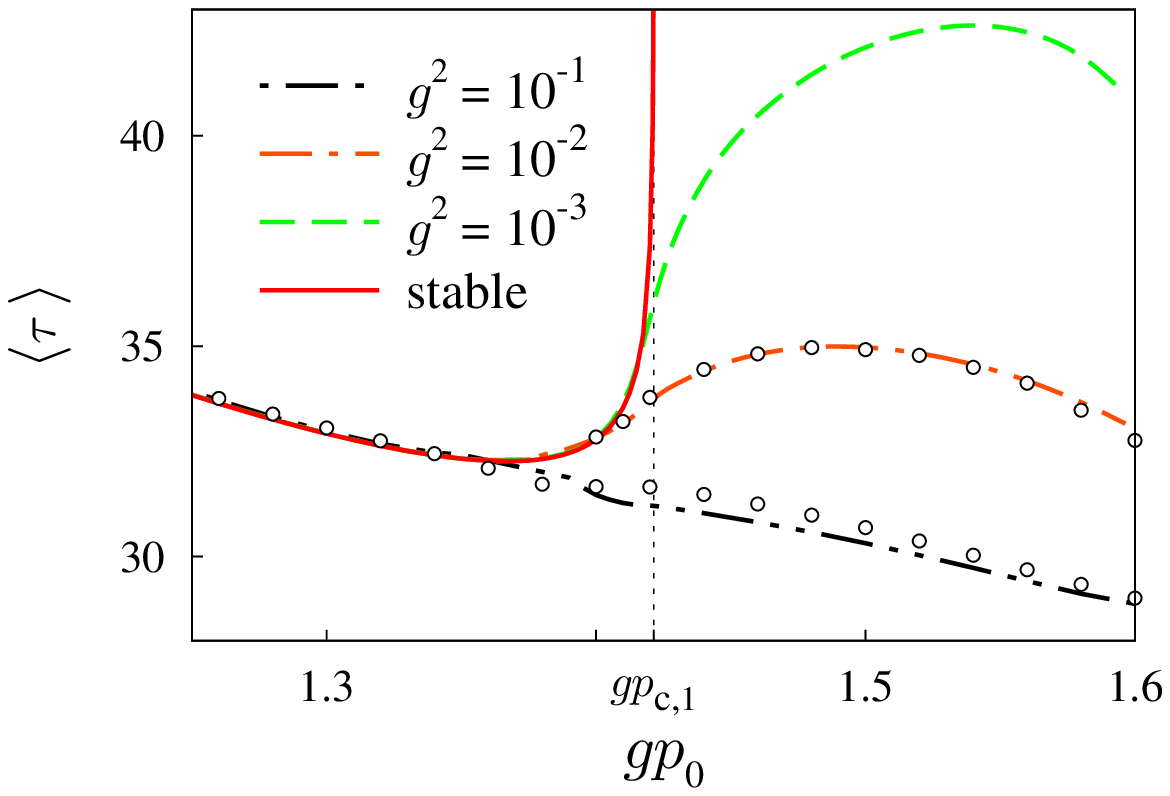}
\includegraphics[width=0.5\textwidth]{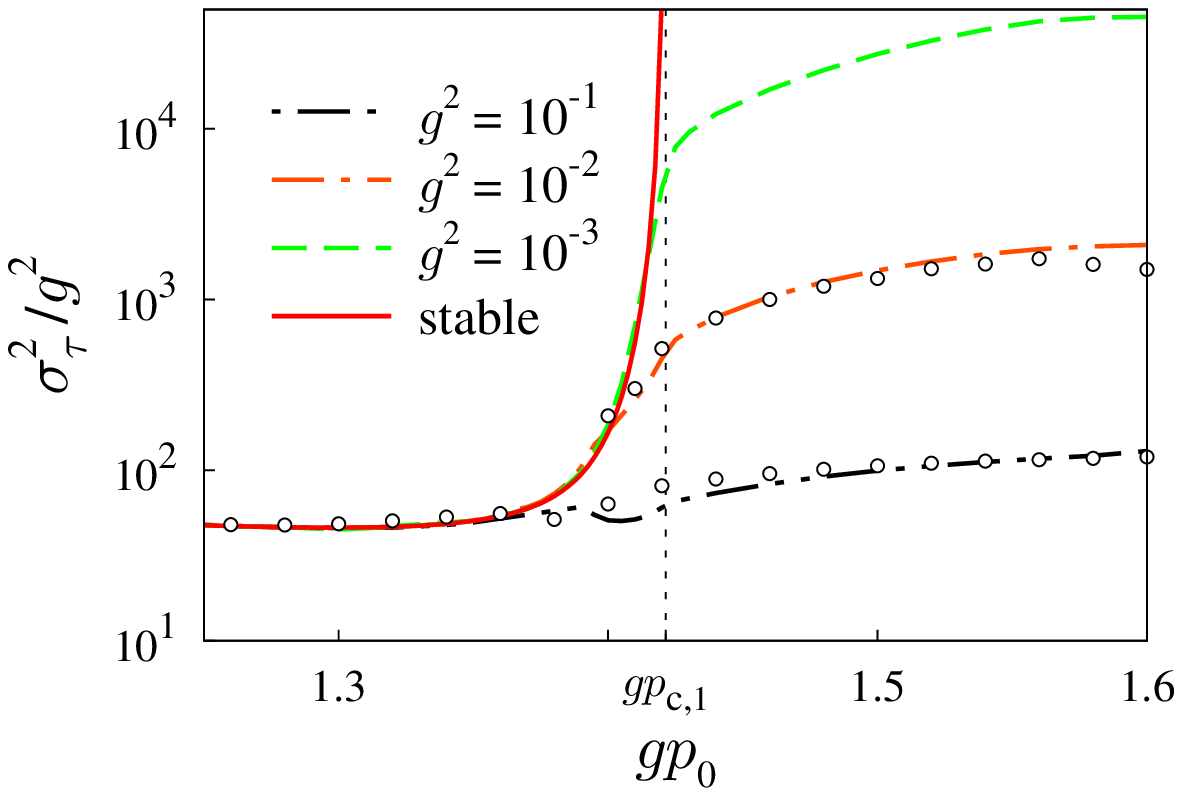}}
\caption{\label{fig:4}(Color online) Mean times of passing $\langle
  \tau \rangle$ and rescaled time dispersions $\sigma_\tau^2/g^2$ as
  functions of $gp_0$. The semiclassical and exact quantum--mechanical
  results are represented by lines and points, respectively. The
  parameters of the graphs are the same as in Figs.~\ref{fig:5}. The
  critical value $gp_{c,1}\approx 1.421$ is shown by the vertical
  dotted line.}
\end{figure}
In Figs.~\ref{fig:4} we plot $\langle \tau\rangle$ and
$\sigma_\tau^2 /g^2$ as functions of mean initial momentum
$p_0$. Dashed lines in these figures represent quantities
computed with the distribution (\ref{eq:75}) at different values 
of $g^2$; they agree with the respective exact quantum--mechanical
data (points). One sees that all graphs in Figs.~\ref{fig:4} coincide
in the region $p_0 < p_{c,1}$ corresponding to stable
trajectories. This is what one expects, since in stable case $\langle
\tau \rangle$ and $\sigma_\tau^2/g^2$ are given by Eqs.~(\ref{eq:81})
and therefore do not depend on  $g^2$. ``Stable''
expressions\footnote{Apparent singularities of ``stable'' graphs at
  $p_0 \to p_{c,1}$  indicate that Eqs.~(\ref{eq:81}) break down in
  the immediate vicinity of $p_{c,1}$.} (\ref{eq:81}) are represented
by the solid lines. Above the critical momentum $p_{c,1}$ the situation in
Figs.~\ref{fig:4} changes drastically: both $\langle  \tau \rangle$ and
$\sigma_\tau^2/g^2$ depend on the semiclassical parameter, so that
small $g^2$ correspond to large mean times of passing and
large time dispersions. 

Finally, let us repeat that the processes of sphaleron--driven
tunneling survive in the limit $\sigma_p^2\to 0$, in contrast to
one--dimensional activation processes of the previous Section. Our
numerical calculations show that, indeed, at arbitrarily small
$\sigma_p^2$ there exists a region of unstable semiclassical 
motions $p_{c,1} < p_0 < p_{c,2}$; the width of this region stays
finite in the limit $\sigma_p^2 \to 0$. 

\subsection{Long--time behavior of traversal--time distribution}
\label{sec:long-time-asympt}
Let us return to the one--dimensional setup of
Sec.~\ref{sec:sphal-driv-tunn} and consider transmission at high
initial momentum $p_0 > \sqrt{2V_0}$. This process
is described by stable over--barrier trajectory $x(t)$ which 
passes the region $x_1 < x < x_2$ in finite time $\tau = \langle \tau
\rangle$.  In the vicinity of $\langle \tau \rangle$ the
distribution $\rho(\tau)$ is Gaussian, see
Sec.~\ref{sec:semiclassical-method-1}. However, we are 
interested in the region $\tau\to +\infty$ where the value of
$\rho(\tau)$ is exponentially small. We are going to show that
in this region $\rho(\tau)$ does not resemble the Gaussian hat
at all; rather, its form is similar to Eq.~(\ref{eq:8}). 

Semiclassically, the asymptotics of $\rho(\tau)$ is elucidated by
studying the properties of the modified trajectories
$x_\epsilon(\tau)$. At large $\tau$ these trajectories linger near the
barrier top, see Eq.~(\ref{eq:11}). One uses considerations of
Sec.~\ref{sec:one-dimension-1} and arrives at the formula
\begin{equation}
\label{eq:23}
\rho_a(\tau) = \rho_\infty \cdot\mathrm{exp}\left\{ -\omega_-
  \tau  - \frac{2\epsilon_0}{g^2\omega_-}\,\mathrm{e}^{-\omega_-
    \tau}\right\}\;,
\end{equation}
cf. Eq.~(\ref{eq:8}). This asymptotics is universal, since parameters
$\omega_-$, $\epsilon_0$ and $\rho_\infty$ disappear after appropriate
shifts and rescalings of $\tau$ and $\rho$. 

Let us point out one important difference between Eqs.~(\ref{eq:23}) and
(\ref{eq:8}). 
In Eq.~(\ref{eq:23}) we consider the case of over--barrier transmissions
when $\rho(\tau)$ reaches its maximum at finite $\tau = \langle \tau
\rangle$. As one goes away from the maximum, the 
exponent $F_\epsilon$ in Eq.~(\ref{eq:75})
grows. Since $\tau$--derivative of this exponent is equal to
$-2\epsilon$, the parameter $\epsilon$ is negative at $\tau > \langle \tau
\rangle$; thus, $\epsilon_0 < 0$
by Eq.~(\ref{eq:12}). One concludes that the asymptotics (\ref{eq:23}) 
monotonically decreases with $\tau$ and $\rho_\infty$ is
exponentially small.

\begin{figure}
\centerline{\includegraphics[width=0.6\textwidth]{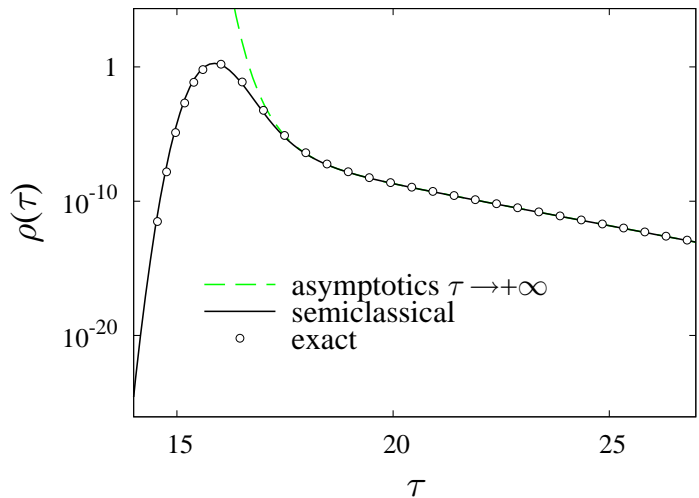}}
\caption{\label{fig:8}(Color online) Distribution $\rho(\tau)$ plotted
  for over--barrier transitions in the model~(\ref{eq:50}); 
  $g^2 = 0.002$, $p_0 = 1.45/g > \sqrt{2V_0}$,
  $\sigma_p^2 = 0.02$, $gx_1 = -gx_2 = 10$. }
\end{figure}
In Fig.~\ref{fig:8} we compare Eq.~(\ref{eq:23}) (dashed
line) with  full semiclassical distribution~(\ref{eq:75})  (solid
line) and exact quantum mechanical results (points). One observes 
coincidence at $\tau \gg \langle \tau \rangle$. 

\section{Summary and remarks}
\label{sec:concluding-remarks}
The qualitative result of this paper is summarized
as follows: quantum transitions take long times if they are described
by unstable semiclassical trajectories. To understand how long this
``long'' is, we introduced the semiclassical parameter $g^2$
in Eq.~(\ref{eq:27}). Then, unstable semiclassical dynamics can be
identified in the limit $g^2\to 0$ where one obtains logarithmically
large mean times of transitions and large time dispersions. This
behavior is drastically different from that in the case of stable
trajectories.

We defined traversal time from time--of--arrival measurements
by the asymptotically distant detector. One may be unsatisfied with
this definition and use another one, see e.g.
Refs.~\cite{Bohm:1951,Smith:1959,Baz:1967,Buttiker:1982,Sokolovski:2008}.
However, the qualitative 
conclusion about large traversal times should be valid for any
reasonable traversal--time definition, since large time delay
comes from the almost--classical motion in the vicinity of the 
unstable mediator orbit (sphaleron or the barrier top).

The starting point of our analysis was general semiclassical expression
(\ref{eq:75}) for the traversal--time distribution $\rho(\tau)$. In
Sec.~\ref{sec:results} we applied this expression to three
processes related to unstable semiclassical dynamics: one--dimensional
activation transitions, multidimensional processes of
sphaleron--driven tunneling and over--barrier transmissions with large
delays. The results were obtained analytically in
one dimension, Eqs.~(\ref{eq:8}) and (\ref{eq:23}), and
numerically in the case of sphaleron--driven tunneling. We
observed the following feature due to semiclassical
instabilities: $\rho(\tau)$ is highly asymmetric and contains long 
exponential tail stretched toward  $\tau \to
+\infty$. This distribution is  
drastically different from Eq.~(\ref{eq:80})
obtained in the case of stable trajectories. 

Note that our results are somewhat different from those of
Ref.~\cite{Takahashi:2006} where {\it power--law} dependence of 
$|\Psi(\tau,x_2)|^2$ on $\tau$ was reported. This dependence should be
compared with the {\it exponential} tail of $\rho(\tau)$ in
Fig.~\ref{fig:5}b. It seems that the difference can be explained in
the following way. In the model of Ref.~\cite{Takahashi:2006} many
semiclassical trajectories contribute substantially into
$\Psi(\tau,x_2)$; one assumes that the number of these
contributions grows with time $\tau$. On the other hand, only one
trajectory is relevant in our model (\ref{eq:58}). Summation over
large number of semiclassical contributions may be responsible for the
modification of large--$\tau$ behavior of wave function in the model
of Ref.~\cite{Takahashi:2006} as compared to our case. 

Unusual form of $\rho(\tau)$ and related scalings of $\langle \tau
\rangle$, $\sigma_\tau^2$ with $g^2$ can be used for experimental
identification of sphaleron--driven tunneling and one--dimensional
activation transitions.

\paragraph*{Acknowledgments.} We are indebted to S.V.~Demidov,
V.A.~Rubakov and S.M.~Sibiryakov for useful discussions and
criticism. This work is supported in part by the grants of the
President of Russian Federation NS-1616.2008.2, MK-1712.2008.2
(D.L.), Russian Science Support Foundation (A.P.), ``Dynasty''
Foundation (awarded by the Scientific board of ICPFM) and 
RFBR grant 08-02-00768-a. Numerical calculations  were performed on
the Computational cluster of Theoretical division of INR RAS.

\appendix

\section{Semiclassical transmission probability}
\label{sec:standard-method}
Let us evaluate semiclassically the total probability 
${\cal P}_\infty$. For definiteness we consider two--dimensional
model of Sec.~\ref{sec:sphal-driv-tunn}; semiclassical calculations in
one dimension are performed in the same way (see comments in
Appendix~\ref{sec:1d}).

One starts with the path integral for the final state,
Eq.~(\ref{eq:60}). The in--state $\Psi_i$ entering this integral
is described in Sec.~\ref{sec:sphal-driv-tunn}. In the semiclassical
approximation one writes,
\begin{equation}
\label{eq:61}
\Psi_i({\bf x}_i) = \frac{\sqrt{2\omega\sigma_p}}
{(2\pi)^{3/4}\sqrt{p_y(y_i)}} \cdot \mathrm{e}^{iB(x_i,\, y_i) + i\pi/4 }\;,
\end{equation}
where $p_y(y_i) = \sqrt{2E_y - \omega^2 y^2}$ is $y$--component of
the initial momentum, and 
\begin{equation}
\label{eq:62}
B(x_i,\, y_i) = i\sigma_p^2 (x_i - x_1)^2  + p_0 x_i +
\int_{\sqrt{2E_y}/\omega}^{y_i} p_y(y') dy'\;,
\end{equation}
is in--state correction to the classical action.

Note that the integrand in Eq.~(\ref{eq:60}) contains fast--oscillating
exponent  $\exp[i(S+B)]$, where $B$ comes from $\Psi_i$. Following the 
prescription of the saddle--point method, one finds the trajectory ${\bf
  x}(t)$ extremizing  the functional $S+B$. In particular, ${\bf
  x}(t)$ should satisfy the classical equations of motion and
Eqs.~(\ref{eq:18}). After the saddle--point
integration one obtains, 
\begin{equation}
\label{eq:64}
\Psi(\tau,{\bf x}_f) = D_1^{-1/2} \cdot \mathrm{e}^{i(S+B) + i\pi/4}\;,
\end{equation}
where all prefactors and saddle--point determinants are collected in
$D_1$; we evaluate them below. We substitute Eq.~(\ref{eq:64}) 
into the probability formula~(\ref{eq:55}) and compute 
the saddle--point integral over ${\bf x}_f$. In this way we derive
the conditions (\ref{eq:19}) and probability expression
(\ref{eq:59}). [Note that $\tau$  is sent to infinity in the final
formulas.] The suppression exponent of the probability is given by the
value of the functional
\begin{equation}
\label{eq:65}
F = 2 g^2\,\mathrm{Im} (S + B)
\end{equation}
on the trajectory ${\bf x}(t)$. The probability prefactor 
\begin{equation}
\label{eq:66}
A = \frac{2\pi}{|D_1|\sqrt{D_2}}
\end{equation}
involves additional determinant $D_2$ due to the saddle--point
integration in Eq.~(\ref{eq:55}). 

Before proceeding with the factors $D_1$ and $D_2$, we preview
the result for the probability prefactor $A$. One considers the set of
perturbations  $\delta {\bf x}(t)$ satisfying the linearized equations
of motion
\begin{equation}
\label{eq:68}
 \delta \ddot{\bf x} + \hat{U}''({\bf x}(t)) \delta {\bf x} = 0
\end{equation}
in the background of complex trajectory ${\bf x}(t)$. The basis in
this set is formed by the ``momentum'' and ``coordinate''
perturbations $\boldsymbol{\psi}^{(j)}(t)$ and
$\boldsymbol{\chi}^{(j)}(t)$, $j=1,2$. We assume that the basal
perturbations satisfy two conditions: (i) they are normalized by
$$
\Omega(\boldsymbol{\psi}^{(j)},\, \boldsymbol{\psi}^{(k)}) =
\Omega(\boldsymbol{\chi}^{(j)},\,
\boldsymbol{\chi}^{(k)}) = 0\;,\qquad\qquad 
\Omega(\boldsymbol{\psi}^{(j)},\, \boldsymbol{\chi}^{(k)}) =
\delta^{jk}\;,
$$
where $\Omega = d{\bf p} \wedge d{\bf x}$ is the canonical symplectic
form; (ii) they are real in the out--region. Note that both of
the above conditions can be imposed in the out--region and therefore
constitute final Cauchy data for the perturbations
$\boldsymbol{\psi}^{(j)}(t)$ and $\boldsymbol{\chi}^{(j)}(t)$.
Note also that the basal perturbations are not completely specified by
(i), (ii); in practice one fixes the freedom by supplying some
additional data.

One introduces the functionals
\begin{equation}
\label{eq:38}
\delta I_1[\delta {\bf x}] = \delta \dot{x}_i - 2i\sigma_p^2 \,\delta
x_i \;,\qquad \qquad \delta I_2[\delta {\bf x}]  = \dot{y}_i \delta
\dot{y}_i + \omega^2 y_i \delta y_i
\end{equation}
which measure changes of initial conditions (\ref{eq:18}) caused by
the perturbations $\delta {\bf x}(t)$. Changes due to the basal
perturbations form two matrices,
\begin{equation}
\label{eq:63}
\Pi_{jk} = \delta I_j[\boldsymbol{\psi}^{(k)}]\;, \qquad \qquad 
\Theta_{jk} = \delta I_j[\boldsymbol{\chi}^{(k)}]\;.
\end{equation}
The probability prefactor is written in terms of these matrices,
\begin{equation}
\label{eq:67}
A = \frac{2\omega \sigma_p}{\sqrt{2\pi \cdot\det [ i\Theta
    \Pi^+ - i \Pi \Theta^+]}}\;.
\end{equation}
This expression is valid for any set of basal
perturbations $\boldsymbol{\psi}^{(j)}$, $\boldsymbol{\chi}^{(j)}$
satisfying (i), (ii).

Semiclassical calculation of transmission probability ${\cal
  P}_{\infty}$ is summarized as follows. One finds the complex
trajectory ${\bf x}(t)$ from the classical equations of
motion and boundary conditions (\ref{eq:18}), (\ref{eq:19}). 
The exponent $F$ is given by the value of the functional
(\ref{eq:65}) on this trajectory. One also constructs four
linear perturbations $\boldsymbol{\psi}^{(j)}$,
$\boldsymbol{\chi}^{(j)}$ by solving Eq.~(\ref{eq:68}) with the Cauchy
data (i), (ii) in the out--region. Prefactor $A$ is computed
by Eq.~(\ref{eq:67}); it involves changes of initial
conditions due to the perturbations $\boldsymbol{\psi}^{(j)}$ and
$\boldsymbol{\chi}^{(j)}$. Given $A$ and $F$, one finds the
probability ${\cal P}_\infty$ from Eq.~(\ref{eq:59}).

We derive Eq.~(\ref{eq:67}) in two steps. First, we completely specify
the basal perturbations $\boldsymbol{\psi}^{(j)}(t)$,
$\boldsymbol{\chi}^{(j)} (t)$ and find $A$ in terms of these
perturbations. Second, we prove that Eq.~(\ref{eq:67}) is 
basis--independent once the conditions (i), (ii) are met.

One computes $D_1$ by collecting the prefactors in  Eq.~(\ref{eq:60}),
\begin{align}
\notag
D_1 &= (2\pi)^{3/2}\frac{\dot{y}_i}{2\omega\sigma_p} \times (2\pi i)^2
 \left[ \det \frac{\partial \dot{\bf x}_f}{\partial {\bf x}_i}
\right]^{-1} \times \frac{1}{(2\pi i)^{2}} \det \frac{\partial^2
  (S+B)}{\partial {\bf x}_i^2} 
\\ &\label{eq:69}
= (2\pi)^{3/2} \frac{\dot{y}_i}{2\omega \sigma_p} \det
\frac{\partial^2 (S+B)}{{\partial \dot{\bf x}_f}\partial{\bf x}_i}
\;,
\end{align}
where ${\bf x}_f$ is kept constant in the differentiations. 
Three factors in the first line come from the initial state
(\ref{eq:61}) and saddle--point integrals over\footnote{We performed
  integration over ${\bf x}(t)$ using the semiclassical Van Vleck
  formula~\cite{VanVleck}.} ${\bf x}(t)$ and ${\bf x}_i$,
respectively. Evaluating explicitly the derivatives of $S$ and $B$,
one obtains,
\begin{equation}
\label{eq:70}
D_1 = \frac{(2\pi)^{3/2}}{2\omega\sigma_p} \det \left[
  \begin{pmatrix} 1& 0 \\ 0 &\dot{y}_i\end{pmatrix} \frac{\partial
    \dot{\bf x}_i}{\partial \dot{\bf x}_f} + \begin{pmatrix} -
    2i\sigma_p^2 & 0 \\  0 & \omega^2 y_i \end{pmatrix}
  \frac{\partial {\bf x}_i}{\partial\dot{\bf x}_f}\right]
\Bigg|_{{\bf x}_f = \mathrm{const}}\;,
\end{equation}
where we wrote $\dot{y}_i$ as a determinant of diagonal matrix and
performed matrix multiplication. 
We choose the basal perturbations $\boldsymbol{\psi}^{(1)}(t) =
\partial {\bf x}(t)/\partial \dot{x}_f$, $\boldsymbol{\psi}^{(2)}(t) =
\partial {\bf   x}(t)/\partial \dot{y}_f$, where the derivatives are
taken at fixed ${\bf x}_f$. Clearly, these perturbations satisfy the 
linearized equations~(\ref{eq:68}). Their final Cauchy data are 
\begin{equation}
\label{eq:76}
\boldsymbol{\psi}^{(1)}(\tau) = \boldsymbol{\psi}^{(2)} (\tau) = 0\;,
\qquad \qquad 
\boldsymbol{\dot{\psi}}^{(1)}(\tau) =
\begin{pmatrix}1\\0\end{pmatrix}\;, \qquad\qquad
\boldsymbol{\dot{\psi}}^{(2)}(\tau) =
\begin{pmatrix}0\\1\end{pmatrix}\;. 
\end{equation}
In terms of $\boldsymbol{\psi}$--perturbations 
\begin{equation}
\notag
D_1 =\frac{(2\pi)^{3/2}}{2\omega\sigma_p} \cdot\det \Pi\;,
\end{equation} 
where $\Pi$ is $2\times 2$ matrix defined in Eq.~(\ref{eq:63}).  

Determinant $D_2$ comes from the integration in Eq.~(\ref{eq:55}),
\begin{equation}
\label{eq:71}
D_2 =  \det \left[ - i \frac{\partial
    \dot{\bf x}_f}{\partial {\bf x}_f} + \mathrm{h.c} \right]\;,
\end{equation}
where the derivatives are taken in the subclass of trajectories
satisfying Eqs.~(\ref{eq:18}). Let us define two remaining basal
perturbations as $\boldsymbol{\chi}^{(1)}(t) = \partial {\bf
  x}(t)/\partial x_f$, $\boldsymbol{\chi}^{(2)}(t) = \partial {\bf
  x}(t)/\partial y_f$, where $\dot{\bf x}_f$ is kept fixed in the
differentiations. These perturbations satisfy 
\begin{equation}
\label{eq:73}
\boldsymbol{\chi}^{(1)}(\tau) = \begin{pmatrix} 1\\0 \end{pmatrix} \;,
  \qquad \qquad 
\boldsymbol{\chi}^{(2)}(\tau) = \begin{pmatrix} 0\\1 \end{pmatrix} \;,
\qquad \qquad
\boldsymbol{\dot{\chi}}^{(1)}(\tau) =
\boldsymbol{\dot{\chi}}^{(2)}(\tau) = 0\;.
\end{equation}
Note that the matrix in $D_2$ is not directly related to
$\boldsymbol{\chi}$--perturbations, since the differentiations in
Eq.~(\ref{eq:71}) are performed with Eqs.~(\ref{eq:18}) kept
fixed. One introduces two auxiliary perturbations 
\begin{equation}
\label{eq:74}
\boldsymbol{\kappa}^{(1)}(t) = \frac{\partial {\bf x}(t)}{\partial x_f}
\Bigg|_{\mathrm{Eqs.~(\ref{eq:18})}}\;,\qquad\qquad
\boldsymbol{\kappa}^{(2)}(t) = \frac{\partial {\bf x}(t)}{\partial y_f}
\Bigg|_{\mathrm{Eqs.~(\ref{eq:18})}}
\end{equation}
which enter Eq.~(\ref{eq:71}). They can be decomposed in
the $\boldsymbol{\psi}$, $\boldsymbol{\chi}$--basis,
\begin{equation}
\label{eq:77}
\boldsymbol{\kappa}^{(j)} = \sum_{k=1}^2 A_{jk}\, \boldsymbol{\psi}^{(k)} +
\boldsymbol{\chi}^{(j)} \;,
\end{equation}
where we took into account Eqs.~(\ref{eq:76}),
(\ref{eq:73}). Since $\boldsymbol{\kappa}$--perturbations do not
change the initial conditions, one writes
$\delta I_j[\boldsymbol{\kappa}^{(k)}] = 0$. Substituting
representation (\ref{eq:77}) into these equations, one obtains 
$$
A^T = - \Pi^{-1} \Theta\;.
$$ We finally write $D_2$ in terms of the perturbations
$\boldsymbol{\kappa}^{(j)}$ and use Eq.~(\ref{eq:77}). We find,
\begin{equation}
\notag
D_2 = |\det \Pi|^{-2}\cdot \det[i \Theta \Pi^+ - i \Pi \Theta^+ ]\;.
\end{equation}
Using expressions for $D_1$, $D_2$ and Eq.~(\ref{eq:66}), one
gets Eq.~(\ref{eq:67}). 

We finish this Appendix by proving that any set of basal perturbations
with properties (i), (ii) can be used in Eq.~(\ref{eq:67}).  Note
first that $\boldsymbol{\psi}^{(j)}$, $\boldsymbol{\chi}^{(j)}$
introduced in Eqs.~(\ref{eq:76}), (\ref{eq:73}) do satisfy these
properties. Denote by ${\boldsymbol{\tilde{\psi}}}^{(i)}$, 
${\boldsymbol{\tilde{\chi}}}^{(i)}$ some other set which also passes (i),
(ii). One decomposes new perturbations in the old basis,
substitutes them into Eq.~(\ref{eq:67}) and uses (i), (ii). In this
way one proves that the value of the prefactor (\ref{eq:67}) does not
depend on the choice for basal perturbations once they are real in the
out--region and correctly normalized.

\section{Expression for $\rho(\tau)$}
\label{sec:modification} 
Here we give details of the semiclassical evaluation of $\rho(\tau)$
which was outlined in Sec.~\ref{sec:semiclassical-method-1}. We start 
by computing the integral in brackets in Eq.~(\ref{eq:28}). The result
is (cf. Eq.~(\ref{eq:64}))
\begin{equation}
\label{eq:5}
\Psi(\tau,{\bf x}_f) = \int_0^\tau d\tau' \int_{i\infty}^{-i\infty}
\frac{id\epsilon}{2\pi g^2} \;
D_{1,\epsilon}^{-1/2} \cdot\mathrm{e}^{i(S_{\epsilon}[{\bf x}_{\epsilon}]
  + B[{\bf x}_{\epsilon}]) + i\pi/4}\;, 
\end{equation}
where $S_\epsilon = S + i\epsilon(T_{int} - \tau')/g^2$ is the
modified action. By subscripts $\epsilon$ here we mark the
quantities computed with the modified trajectory ${\bf x}_\epsilon(t)$
and modified action 
$S_\epsilon$. Recall that ${\bf x}_\epsilon(t)$ extremizes
$S_\epsilon$, arrives at ${\bf x}_{\epsilon}(\tau) = {\bf x}_f$ and
satisfies the initial conditions (\ref{eq:18}).  

Now, we evaluate the integral over $\epsilon$ in
Eq.~(\ref{eq:5}). Since ${\bf x}_\epsilon(t)$ extremizes
$S_\epsilon+B$, one writes,
\begin{equation}
\label{eq:32}
\frac{d}{d\epsilon} (S_\epsilon + B) = 
\frac{d{\bf x}_\epsilon}{d\epsilon} \cdot 
  \frac{\delta}{\delta {\bf x}_\epsilon} (S_\epsilon + B)  
  + \frac{\partial}{\partial \epsilon}
(S_\epsilon + B) = i(T_{int}-\tau')/g^2\;.
\end{equation}
Thus, the exponent in Eq.~(\ref{eq:5}) is extremal with respect to
$\epsilon$ when
$T_{int}[{\bf x}_\epsilon] = \tau'$. Performing the saddle--point 
integration over $\epsilon$, one obtains,
\begin{equation}
\label{eq:29}
\Psi(\tau,{\bf x}_f) = \int_0^\tau \frac{d\tau'}{g\sqrt{2\pi
    D_{1,\epsilon}}} \sqrt{-\frac{d\epsilon}{d\tau'}}
\cdot\mathrm{e}^{i(S_{\epsilon} + B) + i\pi/4}\;.
\end{equation}
Additional prefactor in this expression arises due to fixation of
$T_{int}$.

Next, we substitute $\Psi_f$ into the probability
formula~(\ref{eq:55}). Note that ${\cal P}_\tau$ involves, besides
$\Psi(\tau,{\bf x}_f)$, its complex conjugate. Starting with the path
integral for the latter, one shows that expression for 
$\Psi^*(\tau,{\bf x}_f)$ is obtained from Eq.~(\ref{eq:29}) by
changing the signs of the exponent and $\epsilon$. We write,
\begin{equation}
\label{eq:30}
{\cal P}_\tau = \int d{\bf x}_f
\int_0^{\tau} \frac{d\tau' d\bar{\tau}} 
              {2\pi g^2 \sqrt{D_{1,\epsilon} D_{1,-\bar{\epsilon}}}} 
  \sqrt{\frac{d\epsilon}{d\tau'}
        \frac{d\bar{\epsilon}} {d\bar{\tau}}} \cdot
 \mathrm{e}^{i(S_\epsilon[{\bf x}_\epsilon] + B[{\bf
 x}_\epsilon]) -
 i(S_{-\bar{\epsilon}}[{\bf x}_{-\bar{\epsilon}}] + B[{\bf
 x}_{-\bar{\epsilon}}])} \;,
\end{equation}
where $\bar{\tau}$ and $\bar{\epsilon}$ come from $\Psi^*(\tau,{\bf
  x}_f)$; they are related by $T_{int}[{\bf x}_{-\bar{\epsilon}}] =
  \bar{\tau}$.

 One sees that the probability formula involves integrals over {\it two}
traversal times $\tau'$ and $\bar{\tau}$ spent by the respective
trajectories in the region $x<x_2$. We introduce new variables
$\tau_+ = (\tau' + \bar{\tau})/2$ and $\tau_- = \tau' - \bar{\tau}$
and perform the integral over $\tau_-$ in the saddle--point
approximation. This leads to the condition
$\epsilon=\bar{\epsilon}$. Note that after the integration the
value of $\epsilon$ is related to $\tau_+$ by the
implicit relation
\begin{equation}
\label{eq:31}
T_{int}[{\bf x}_\epsilon] + T_{int}[{\bf x}_{-\epsilon}] = 2\tau_+\;.
\end{equation}
Assuming that the trajectory ${\bf x}_\epsilon(t)$ is unique, one
proves that the saddle--point value of $\epsilon$ is real. Indeed, it
is straightforward to check that ${\bf x}_\epsilon^* = {\bf
  x}_{-\epsilon^*}$. Thus, for real $\epsilon$ the function
$\tau_+(\epsilon)$ is real, see Eq.~(\ref{eq:31}). The same is
correct for the inverse function $\epsilon(\tau_+)$. Equation
(\ref{eq:31}) and reality of $\epsilon$ lead to the relation 
\begin{equation}
\label{eq:25}
\mathrm{Re}\, T_{int}[{\bf x}_\epsilon] = \tau_+
\end{equation}
which implies that $\tau_+$ is  the real part of the time interval
spent by the modified trajectory in the region between $x_1$ and
$x_2$.

Finally, let us evaluate the integral over ${\bf x}_f$ in
Eq.~(\ref{eq:30}). It is taken in 
the same way as in Appendix~\ref{sec:standard-method}. However, we obtain
conditions at finite $\tau$,
\begin{equation}
\label{eq:9}
 \mathrm{Im}\, \dot{\bf x}_{\epsilon}(\tau) =  \mathrm{Im}\, {\bf
   x}_{\epsilon}(\tau) = 0  
\end{equation}
cf. Eqs.~(\ref{eq:19}).

We arrive at the following representation,
\begin{equation}
\label{eq:20}
{\cal P}_\tau = \int_0^{\tau} \frac{d\tau_+}{\sqrt{\pi g^2}}
\left[ -\frac{d\epsilon}{d{\tau_+}}\right]^{1/2} A_{\epsilon} \cdot
\mathrm{e}^{-F_{\epsilon}/g^2}\;,
\end{equation}
where only the integral over $\tau_+$ is left. Quantities
$F_{\epsilon}$ and $A_{\epsilon}$ in this formula are given by the
same expressions~(\ref{eq:65}) and (\ref{eq:67}), but with the 
modified action (\ref{eq:72}) and modified trajectory ${\bf
  x}_\epsilon(t)$.  Note that now $\tau_+$ enters the modified action
$S_\epsilon$ instead of $\tau'$.

In Refs.~\cite{epsilon} representation similar to Eq.~(\ref{eq:20}) was
used for evaluation of total transmission probability ${\cal
  P}_\infty$. In this case one sends $\tau$ to infinity and performs
the integral over $\tau_+$. We follow another path and calculate
$\rho(\tau)$ by differentiating Eq.~(\ref{eq:20}) with respect to
$\tau$.

Let us show that the integrand in Eq.~(\ref{eq:20}) does not depend
on the duration of the process $\tau$. Consider first the modified
trajectory ${\bf x}_\epsilon(t)$. Parameter $\tau$ enters
the  boundary value problem for ${\bf x}_\epsilon(t)$ via 
the conditions (\ref{eq:9}) imposed at $t = \tau$. Note, however, that
${\bf x}_\epsilon(\tau)$ lies {\it inside} the detection region
$x>x_2$. Indeed, due to Eq.~(\ref{eq:25}) $\tau_+$ is equal to the time
of motion in the region $x<x_2$; at $\tau_+ < \tau$ 
the trajectory leaves this region  {\it before} $t =
\tau$. From Eqs.~(\ref{eq:72}), (\ref{eq:2}) one finds that the
classical equations of motion are not modified in the detection region
$x>x_2$. Thus, ${\bf x}_\epsilon(t)$ describes real classical
evolution at $t\approx \tau$, and conditions of reality, 
Eqs.~(\ref{eq:9}), can be imposed at any point of this evolution
without affecting the trajectory. One concludes that  ${\bf
  x}_\epsilon(t)$  does not depend on $\tau$. The suppression 
exponent $F_\epsilon$, Eq.~(\ref{eq:65}), is independent of $\tau$ as
well, since it involves the {\it imaginary} part of 
$S_\epsilon$ computed on the modified trajectory. Function
$\epsilon(\tau_+)$ 
is defined in Eq.~(\ref{eq:25}); it does not depend on $\tau$ due to
definition of $T_{int}$, see Eq.~(\ref{eq:2}). 

Now, we change the duration of the process, $\tau\to \tau + \delta \tau$,
and find the respective changes of the basal perturbations
$\boldsymbol{\psi}$, $\boldsymbol{\chi}$. For concreteness we use 
the perturbations defined by Eqs.~(\ref{eq:76}), (\ref{eq:73}). Solving
explicitly Eqs.~(\ref{eq:68}) in the asymptotic region $x>x_2$, we
find that
\begin{equation}
\label{eq:48} 
\begin{pmatrix} 
\boldsymbol{\psi}^{(1)} \\
\boldsymbol{\chi}^{(1)}
\end{pmatrix} 
\to \begin{pmatrix} 1 &
  -\delta \tau \\ 0 &
  1 \end{pmatrix} \begin{pmatrix}\boldsymbol{\psi}^{(1)} \\
  \boldsymbol{\chi}^{(1)}  \end{pmatrix}\;, \qquad
\begin{pmatrix} 
\boldsymbol{\psi}^{(2)} \\
\boldsymbol{\chi}^{(2)}  
\end{pmatrix} 
\to \begin{pmatrix} \cos(\omega\delta \tau) &
  -\sin(\omega \delta \tau)/\omega \\ \omega\sin(\omega\delta
  \tau) & 
  \cos(\omega\delta \tau) \end{pmatrix}
\begin{pmatrix}\boldsymbol{\psi}^{(2)} \\    
  \boldsymbol{\chi}^{(2)}  \end{pmatrix}\;,
\end{equation}
due to $\delta \tau$. Using these transformations, one explicitly
proves that $A_\epsilon$, Eq.~(\ref{eq:67}), does not depend on
$\tau$.

Now, take a look at the probability ${\cal P}_\tau$,
Eq.~(\ref{eq:20}). We have shown that the quantities $F_\epsilon$,
$\epsilon(\tau_+)$, $A_\epsilon$ in the integrand of this
expression do not depend on $\tau$. Thus, the derivative $\rho(\tau)$,
Eq.~(\ref{eq:1}), is simply given by the 
integrand at $\tau_+ = \tau$. In this way one gets the 
semiclassical expression (\ref{eq:75}) of
Sec.~\ref{sec:semiclassical-method-1}; the relation (\ref{eq:78}) is 
obtained from Eq.~(\ref{eq:25}).

Finally, we prove Eq.~(\ref{eq:79}):
\begin{equation}
\label{eq:6}
\frac{dF_\epsilon}{d\tau} = \frac{\partial F_\epsilon}{\partial\tau} = 
-2\epsilon\;,
\end{equation}
since $F_\epsilon$ is extremal with respect to ${\bf x}_\epsilon(t)$
and $\epsilon$.

\section{Semiclassical calculations in one dimension}
\label{sec:1d}
In this Appendix we perform explicit semiclassical calculations
in the model~(\ref{eq:50}). 

For a start, we find the original complex trajectory $x(t)$. General
classical solution with energy 
$E<V_0\equiv 1/g^2$ looks like 
\begin{equation}
\label{eq:16}
\sqrt{\frac{g^2 E}{1-g^2E}} \cdot \mathrm{sh}(gx) = 
\mathrm{ch}\, \left[ g\sqrt{2E}(t - t_0)\right]\;,
\end{equation}
where $E$ and $t_0$ are real by Eqs.~(\ref{eq:19}). One
notes that the solution (\ref{eq:16}), if taken along the real time
axis, describes reflection from the potential barrier; trajectory
with correct asymptotics is obtained by winding\footnote{Euclidean
  parts of this contour correspond to motions under the
  potential barrier.} the time contour around the nearest upper
branching point of the solution. From the initial condition
(\ref{eq:18}) one finds\footnote{Recall that initial conditions are
  imposed in the asymptotic region $x\to -\infty$ where motion is
  linear.}  the energy of the solution,
$E=(p_0+2\pi\sigma_p^2/g)^2/2$. Since $E$ does not exceed the
height of the potential barrier, the value of $p_0$ is smaller than 
$p_{c,1}$, Eq.~(\ref{eq:4}). 

The case $E>V_0$ corresponds to classical over--barrier motions
which trivially produce $E=p_0^2/2$ and therefore $p_0 >
\sqrt{2V_0} = p_{c,2}$. One concludes that transitions in the region
$p_{c,1} < p_0 < p_{c,2}$ are not described by the trajectories with
$E\ne V_0$. The remaining solution
\begin{equation}
\label{eq:14}
\mathrm{sh}(gx) = -\mathrm{exp}\left[ -\sqrt{2} (t - t_0)\right]
\end{equation}
corresponds to $E=V_0=1/g^2$. It starts in the in--region and approaches
the barrier top as $t\to +\infty$. Note that
Eq.~(\ref{eq:14}) automatically passes the final boundary conditions
(\ref{eq:19}). Thus, $t_0$ is complex; one finds it from
Eq.~(\ref{eq:18}). Trajectories (\ref{eq:14}) exist at any
$p_0$. Calculating the suppression exponent (\ref{eq:65}),
one explicitly checks that they are sub--dominant outside the interval
$p_{c,1} < p_0 < p_{c,2}$ and describe transitions through the barrier
if $p_0$ belongs to this interval. 

Note that the fixed--energy transmission
coefficients ${\cal P}_p$ are exactly known in the model
(\ref{eq:50}). Using these coefficients and Eq.~(\ref{eq:51}), one 
explicitly confirms the semiclassical expressions for ${\cal P}_\infty$,
Refs.~\cite{epsilon}, in cases of stable and unstable semiclassical
dynamics.

We proceed by introducing modification. One notices that the
modified potential~(\ref{eq:22}) involves step--function and therefore
has {\it two} analytic continuations, which start from the regions
$x<x_2$ and $x>x_2$. We find the corresponding parts of the modified
trajectory and sew them and their time derivatives at $t=t_2$ where
$x_\epsilon(t_2) = x_2$. Note that the energy $E$ of trajectory is conserved
across the sewing point. The time $t_2$ and energy $E$ are real due to
the final boundary conditions (\ref{eq:19}). Moreover, $t_2 = \tau$ by
Eq.~(\ref{eq:78}).  At $t<t_2$ the modification adds the  
imaginary  constant $-i\epsilon/g^2$ to the potential. Thus, in this
region $x_\epsilon(t)$ is given by Eq.~(\ref{eq:16}) where
one should substitute $E\to E_\epsilon = E + i\epsilon/g^2$. Note that
parameter $t_0$ is {\it complex} after modification: it is found from
$x_\epsilon(\tau) = x_2$. Using this condition and Eq.~(\ref{eq:18}), 
one obtains non--linear complex equation
\begin{equation}
\label{eq:26}
f(E_\epsilon) \equiv \frac{i}{2\sigma_p^2}\left(\sqrt{2E_\epsilon}-p_0\right) +
  x_2-x_1- \sqrt{2E_\epsilon}
    \,\tau  + \frac1g \cdot\mathrm{ln}\, \frac{g^2
    E_\epsilon}{g^2 E_\epsilon-1} = 0
\end{equation}
which relates $E$ and $\epsilon$ to $p_0$ and $\tau$.

Now, we compute $\rho(\tau)$ using the modified quantities $\epsilon$,
$F_\epsilon$, $A_\epsilon$ and Eq.~(\ref{eq:75}). The function 
$\epsilon(\tau)$ is found from Eq.~(\ref{eq:26}). The
modified exponent
$F_\epsilon$ is computed  by substituting
$S_\epsilon$ and $x_\epsilon(t)$ into Eq.~(\ref{eq:65}) and performing
integration. [We disregard the $y$--dependent part of the functional
  $B$, Eq.~(\ref{eq:62}).] The  final ingredient is the probability
prefactor $A_\epsilon$ which is given by 
one--dimensional formula
\begin{equation}
\label{eq:21}
A = \frac{\sigma_p \sqrt{2}}{\sqrt{\mathrm{Im}(\delta
    I_1^*[\chi]\cdot\delta I_1[\psi])}}\;.
\end{equation}
Here by omitting the subscript $\epsilon$ we mean that
Eq.~(\ref{eq:21}) 
can be used both in modified and unmodified cases. Note that
one--dimensional prefactor can be obtained from Eq.~(\ref{eq:67}) by
reducing the size of matrices $\Pi$ and $\Theta$ and omitting the
kinematical factor $\omega/\sqrt{2\pi}$. 

Linear perturbations $\chi_\epsilon$, $\psi_\epsilon$ are calculated
by differentiating the trajectory 
$x_\epsilon(t)$ with respect to parameters, 
$$
\chi_\epsilon(t) = \frac{\dot{x}_\epsilon(t)}{\sqrt{2E}}\;,\qquad \qquad
\psi_\epsilon(t) = \alpha \dot{x}_\epsilon(t) + \beta \frac{\partial
x_\epsilon(t)}{\partial p_0}\;, 
$$
where we have already used the Cauchy data (\ref{eq:73}) for 
$\chi(t)$. Conditions (\ref{eq:76}) give $\beta^{-1} = \partial \sqrt{2
  E}/\partial p_0$, where the derivative is taken at
$\epsilon=\mathrm{const}$. Substituting the perturbations into 
Eq.~(\ref{eq:21}), and using Eq.~(\ref{eq:18}), one arrives at 
the expression 
$$
A_\epsilon=
\left(
\frac{\partial E}{\partial p_0} \Big/
\mathrm{Re}\,\sqrt{2E_\epsilon} \right)^{1/2}
$$
for the probability prefactor.

We substitute $F_\epsilon$ and $A_\epsilon$ into Eq.~(\ref{eq:75}) and
use some algebra. The result is
\begin{equation}
\label{eq:17}
\rho(\tau) = {\cal N}'\cdot
\left|\frac{df(E_\epsilon)}{dE_\epsilon}\right|^{-1} \cdot 
\mathrm{exp} \left\{ -\frac{1}{\sigma_p^2} \left(p_0^2/2-E\right) -
  \frac{2\epsilon \tau}{g^2} +
    \frac{2\sqrt{2}}{g^2}\mathrm{arg} \left( \frac{\sqrt{E_\epsilon}+1/g} 
      {\sqrt{E_\epsilon} - 1/g} \right) \right\}\;,
\end{equation}
where ${\cal N}'$ is normalization factor. Note that 
$f(E_\epsilon)$ is defined in Eq.~(\ref{eq:26}), and the values of
$E$, $\epsilon$ are found from $f(E_\epsilon)=0$.

The expectation value $\langle \tau \rangle$ of
the distribution~(\ref{eq:17}) is plotted in Fig.~\ref{fig:6}b, solid
line.  ``Stable'' and ``unstable'' distributions, Eqs.~(\ref{eq:80})
and (\ref{eq:8}),~(\ref{eq:13}) are obtained by considering
Eqs.~(\ref{eq:26}), (\ref{eq:17}) in the region of small $\epsilon$.

\section{Exact results in two dimensions}
\label{sec:exact-quant-mech}
We compute exact traversal--time distributions in the model
(\ref{eq:58}) using the following numerical procedure. 

First of all, we introduce the basis of stationary eigenfunctions
$\psi_p(x,y)$ satisfying time--independent Schr\"odinger equation
with scattering boundary conditions,
\begin{equation}
\label{eq:35}
\begin{array}{ll}
\psi_p(x,y) \to \psi_{y}(y) \cdot \mathrm{e}^{ipx} + \mbox{reflected
  waves}\;, \qquad&\mbox{as} \qquad x\to -\infty\;,\\
\psi_p(x,y) \to \mbox{outgoing waves}\;, &
\mbox{as}  \qquad x\to +\infty\;,
\end{array}
\end{equation}
where $\psi_y(y)$ is the initial oscillator state with energy $E_y$.
The total energy corresponding to $\psi_p(x,y)$ is $E_p = p^2/2 + E_y$. 

One decomposes the initial wave packet $\Psi_i$
(Sec.~\ref{sec:sphal-driv-tunn}) in the above basis, 
\begin{equation}
\nonumber
\Psi_i(x,y) = \psi_x(x)\cdot \psi_y(y) = \frac{1}{\sqrt{2\pi}}\int
dp \; \psi_x(p) \, \mathrm{e}^{ipx} \cdot \psi_y(y) =
\frac{1}{\sqrt{2\pi}}\int dp\; \psi_x(p) \, \psi_p(x,y)\;, 
\end{equation}
where we performed Fourier transformation and used the
asymptotics~(\ref{eq:35}) of $\psi_p(x,y)$. Then, we propagate the
wave packet from $t=0$ to $t=\tau$,
\begin{equation}
\label{eq:34}
\Psi(\tau,\, x, y) = \frac{1}{\sqrt{2\pi}}\int dp\; \psi_x(p) \,
\psi_p(x,y)\cdot\mathrm{e}^{-iE_p \tau}\;.
\end{equation}
Given representation~(\ref{eq:34}) one computes the total
probability 
current through the line $x=x_2$. The distribution $\rho(\tau)$ is
proportional to this current, see Sec.~\ref{sec:tunn-time-defin}; the
proportionality 
coefficient is obtained from the normalization condition $\int 
\rho(\tau)\,d\tau = 1$.

In practice we computed the exact distributions $\rho(\tau)$
at\footnote{The values of other  parameters are listed in the caption
  of Figs.~\ref{fig:5}.} $g^2 = 
0.1$ and $0.01$. At each value of $g^2$ we generated the set of eigenfunctions
$\psi_p(x,y)$ by solving numerically the stationary Schr\"odinger
equation at several energies\footnote{The energies were uniformly
  distributed in the interval $g^2 E_y<g^2 E_p < 2$  
  with spacing $g^2\Delta E_p = 0.05$ at $g^2 = 0.1$ and in the
  interval $1 < g^2 E_p < 1.5$ with spacing $g^2 \Delta E_p = 0.02$ at
  $g^2=0.01$.} $E_p$.  To this end we used the numerical
method of Ref.~\cite{Bonini:1999kj} (see Ref.~\cite{f90code} for the
Fortran~90 code). Given the eigenfunctions, we constructed the linear
combination (\ref{eq:34}) and calculated the total probability current
through the line $x=x_2$. Computing numerically the normalization
integral, we finally obtained $\rho(\tau)$.

There are two limitations of our numerical procedure; both of them are
relevant for the numerical data at $g^2=0.01$. First, our method of
solving the Schr\"odinger equation fails to produce correct
eigenfunctions\footnote{This happens when the typical values of
  $\psi_p(x_2,y)$ become comparable to the round--off errors.}
at $g^2=0.01$ and $E_p<1$. Accordingly, the exact data at $g^2=0.01$
in Figs.~\ref{fig:4} are limited to the region $gp_0 \geq 1.4$ where the
contributions of low--energy eigenfunctions are
negligible. Second, due to cancellations in the  integral
(\ref{eq:34}) we are not able to compute directly the probability
current at large $\tau$ (see, e.g., Figs.~\ref{fig:5} where the exact 
distributions are obtained at $\tau \lesssim 45$). This prevents us
from normalizing directly the $g^2=0.01$ distributions which fall off
slowly at large $\tau$. Due to this feature we consider
the normalization--independent quantity $\rho(\tau)/\rho_{\max}$ in
Figs.~\ref{fig:5}. The exact data at $g^2=0.01$ in Figs.~\ref{fig:10}
are obtained by  extrapolating the corresponding distributions with
the conjectured behavior (\ref{eq:36}). After extrapolation we
normalize the distributions and compute the values of $\langle \tau
\rangle$ and~$\sigma_\tau^2$.

We stress that the limitations listed above are not present at
$g^2=0.1$; in this case the exact distributions $\rho(\tau)$ can
be computed directly at all values of $g p_0$. 


\end{document}